\documentclass[twocolumn,showpacs,preprintnumbers,amsmath,amssymb,prx]{revtex4-2}
\usepackage[scr=boondoxo,scrscaled=1.05]{mathalfa}
\usepackage{tikz}
\usetikzlibrary{calc,decorations.markings}
\usepackage{epsfig}
\usepackage{graphicx}
\usepackage{rotating}
\usepackage{amssymb}
\usepackage{amsmath}
\usepackage{subcaption}
\usepackage{float}   %needed for option H (non floating figure)
\usepackage[font=small,labelfont=bf,justification=justified,format=plain]{caption}
\usepackage[active]{srcltx}
\usepackage{url}
\usepackage{hyperref}
\usepackage{centernot}
\usepackage{dsfont,bm}
\usepackage{xcolor}
\usepackage{pgfplots}
\usepackage{stackengine}
\usepackage{enumitem}
\usepackage{float}  % no floating pictures
\usepackage[scr=boondoxo,scrscaled=1.05]{mathalfa}
\hypersetup{
	colorlinks   = true, %Colours links instead of ugly boxes
	urlcolor     = blue, %Colour for external hyperlinks
	linkcolor    = blue, %Colour of internal links
	citecolor   = red %Colour of citations
}
\usepackage{tikz}
\usepackage{tkz-euclide}
\usetikzlibrary{calc,decorations.markings}
\usepackage{times,graphicx,xcolor}
\usepackage{delarray,fancybox}
\usepackage{mathdots}
\usepackage{eurosym}
\usepackage{relsize}
\usepackage{soul}
\usepackage{hexbrace}
\colorlet{darkred}{red!55!black}
\colorlet{darkgreen}{green!25!black}
\newcommand{\ti}{t_{\mathfrak{0}}}
\newcommand{\tf}{t_{\mathfrak{f}}}
\begin{document}

	\title{Unraveling-paired dynamical maps can recover the input of quantum channels}

	\author{Brecht Donvil}
	\affiliation{University of Ulm, Institute for Complex Quantum Systems, Albert-Einstein-Allee 11
		89069 Ulm, Germany}
	\email{brecht.donvil@uni-ulm.de}
	\author{Paolo Muratore-Ginanneschi}
	\affiliation{University of Helsinki, Department of Mathematics and Statistics
		P.O. Box 68 FIN-00014, Helsinki, Finland}
	\email{paolo.muratore-ginanneschi@helsinki.fi}

	\begin{abstract}	
		We explore algebraic and dynamical consequences of unraveling  general time-local master equations. We show that the 
		\textquotedblleft influence martingale\textquotedblright, the paramount ingredient of a recently discovered unraveling framework,  
		pairs any time-local master equation with a one parameter family of Lindblad-Gorini-Kossakowski-Sudarshan master equations.  At any instant of time, the variance of the influence martingale provides an upper bound on the Hilbert-Schmidt distance between solutions of paired master equations. 
		Finding the lowest upper bound on the variance of the influence martingale yields an explicit criterion of \textquotedblleft optimal pairing\textquotedblright.  The criterion independently retrieves the measure of isotropic noise necessary for the structural physical approximation of the flow the time-local master equation with a completely positive flow. The optimal pairing also allows us to invoke a general result on linear  maps on operators (the \textquotedblleft commutant representation\textquotedblright) to embed the flow of a general master equation in the off-diagonal corner of a completely positive map which in turn solves a time-local master equation that we explicitly determine.
		We use the embedding to reverse a completely positive evolution, a quantum channel, to its initial condition thereby providing a protocol to preserve quantum memory against decoherence.  We thus arrive at a model of continuous time error correction by a quantum channel.
	\end{abstract}
	
	\pacs{03.65.Yz, 42.50.Lc}
	
	%\keywords{}
	\maketitle

	\section{Introduction}
	A consequential result of stochastic analysis is the representation of solutions of semi-linear differential equations as Monte Carlo averages over random paths via Feynman-Kac type formulae \cite{DeMo2004}. The result has ubiquitous applications because both in classical and quantum physics Monte Carlo methods often provide the only viable integration strategy when the number of involved degrees of freedom becomes large \cite{ParG1988}. 
	
	In the theory of open quantum systems, the representation of state operators as Monte Carlo averages over pure state operators generated by classical stochastic processes is called unraveling in quantum trajectories \cite{CarH1993}.  
	Different forms of unravelings have been discovered in connection to quantum measurement \cite{SrDa1981,HuPa1984,BaBe1991,GaPaZo1992,BaHo1995,MeStLu2020}, possibly with delay and retrodiction \cite{DiGiSt1998,WiGa2008,DioL2008b}, to microscopic state reduction \cite{GisN1984,GhRiWe1985,BaGh2003,PerI2003}, in addition to the scope of developing efficient computational algorithms in high dimensional Hilbert spaces \cite{DaCaMo1992,StGr2002,PiMaHaSu2008,SuEiSt2014}.  We refer to \cite{WiMi2009,WeMuKiScRoSi2016} for an overview of further recent applications. 
	
	A theoretical result on the dynamics of open system state operators on finite dimensional Hilbert spaces supports the generic existence of unravelings. Namely, it is possible to prove that solutions of the Nakajima-Zwanzig integro-differential equation \cite{ZwaR2001} also satisfy a time-local master equation almost everywhere in time \cite{VstV1973,GrTaHa1977,vaWoLe1995,AnCrHa2007,ChKo2010}.
	The proof requires the existence of a pseudo-inverse of the Nakajima-Zwanzig solution map and is non-constructive. Nonetheless, theoretical considerations permit to infer a unique and thus canonical form of the time-local master equation \cite{HaCrLiAn2014} but not its explicit expression given a microscopic dynamics.  Explicit expressions are typically obtained by means of asymptotic expansions around scaling limits.
	
	The most well known, and mathematically rigorous derivation \cite{DavE1976} relies on the weak-coupling scaling limit, and leads to the master equation derived by Lindblad \cite{LinG1976}, Gorini, Kossakowski and Sudarshan \cite{GoKoSu1976} from the necessary and sufficient conditions to generate a completely positive linear \emph{flow} \cite{ArnV2006} on operators. Beyond weak coupling,	time-convolutionless perturbation theory \cite{ShTaHa1977,HaShSh1977,ChSh1979,BrPe2002,SmVa2010,KiWiRa2015,FeZiBl2021,NeBrWe2021} yields a systematic way to explicitly derive  time-local master equations.
	
	In the presence only of regularity assumptions,  a time-local master equation generates a completely bounded \cite{PauV2003} flow i.e. an infinitely divisible family  \cite{WoCi2008} of linear maps on operators. Completely bounded flows may describe positive and even completely positive dynamical maps when restricted to subsets of initial conditions \cite{SuSh2003}. %From the foundational point of view, the legitimacy of non-completely positive dynamical maps is contentious see e.g. \cite{DoLi2016,ScRiSp2019} although the phenomenological use is  accepted \cite{HaSt2020}. 
	
	Very recently \cite{DoMG2022}, we introduced a framework for unraveling (solutions of) the general time-local and sufficiently regular master equation.
	The conceptual point of the unraveling is to enforce the Wittstock-Paulsen canonical form of a completely bounded map \cite{WitG1981,PauV1982}. To achieve this goal, we included in the Monte-Carlo average pure states compatible with the most general, non-linear with respect to the initial data, physically admissible Kraus-type form \cite{ToKwOhChMa2004,SaSaGoFe2004}. A mean preserving martingale of the stochastic process generating the dynamics - the \emph{influence martingale}, as we called it - attributes the proper weight and sign to each pure state in order to recover a completely bounded map, on average.
	
	In this paper, we explore algebraic and dynamical consequences of unravelings based on the influence martingale and obtain three main results.
	
	We start from the universal canonical form of the time-local master equation whose properties we recall in section~\ref{sec:cb}. We prove that it is always possible to construct the unraveling using only pure-states whose evolution law corresponds to the completely positive Kraus form \cite{BrPe2002}. In other words, we restrict the generation of the statistics to 
	pure states	whose average in the absence of the influence martingale specifies a completely positive flow.
	In such a formulation of the unraveling, the influence martingale pairs a completely bounded master equation to a one-parameter family of completely positive master equations (section~\ref{sec:unraveling}). It is thus possible to introduce a notion of optimality of the pairing (section~\ref{sec:variance}) based on the minimization of the Hilbert-Schmidt distance of a snapshot of the solution of the completely bounded master equation from that of an element of the completely positive family. Namely, the variance of the influence martingale always provides an upper bound on the 
	Hilbert-Schmidt squared distance. The dynamics of the influence martingale is always enslaved to that of random pure-state operators whose contribution to the Monte Carlo average it weighs. Nevertheless, it is possible to find a lowest upper bound to the variance of the influence martingale universal with respect to the state of the stochastic process and thus providing an \textquotedblleft optimal pairing\textquotedblright\ criterion. Remarkably, the optimal criterion implies that  the growth rate of the influence martingale must be proportional to the minimum rate of quantum isotropic noise that must be added to a completely bounded flow over an infinitesimal time step to recover a completely positive evolution \cite{WoEiCuCi2008,HaCrLiAn2014}. Quantum isotropic noise is described the maximally mixed state generated by the depolarizing map. Adding a depolarizing channel to a positive flow serves the purpose of engineering entanglement witness operators suited to laboratory implementation. The procedure is referred to as the structural physical approximation \cite{HoEk2002}. Thus, the optimal pairing criterion recovers physical information specifying a well known indicator of the deviation of a flow on operators from a completely positive evolution \cite{RiHuPl2014}.  Altogether, these results constitute the first main contribution of the paper.

	In section~\ref{sec:embedding}, we derive an algebraic property of the optimal pairing. Namely, optimally paired completely positive and completely bounded flows, the latter rescaled by a universal factor, specify the blocks of a two parameter family of completely positive maps on operators acting on the extended Hilbert space $ \mathbb{C}^{2}\otimes \mathcal{H} $. The embedding family is in turn solution of a Lindblad-Gorini-Kossakowski-Sudarshan, or completely positive, master equation which we explicitly determine.
	We thus introduce a notion of optimal embedding of a completely bounded map which is a second main result of this work. The existence of the embedding is guaranteed by a corollary to the proof of existence of the \textquotedblleft commutant\textquotedblright\ representation of completely bounded maps \cite{PaYuSu1985}. We also contrast (section~\ref{sec:norm}) the optimization we introduce with the one interwoven with the definition of the \textquotedblleft triple norm\textquotedblright\ of a completely bounded map \cite{PaYuSu1985}.
	
	The optimal embedding paves the way to a protocol to recover the initial state of an open quantum system by means of a measurement on another completely positive evolution. Namely, in section~\ref{sec:recqc} we consider the dynamics of the inverse flow of 
	a completely positive evolution. Next, we use the results of section~\ref{sec:embedding} to embed the inverse flow into a completely positive flow on operators on $\mathbb{C}^{2}\otimes \mathcal{H} $. Physically, the embedding corresponds to coupling a completely positive evolution in the Hilbert space of the original system to an ancillary qubit. If the input to the quantum channel on $\mathbb{C}^{2}\otimes \mathcal{H} $ is the tensor product of an adapted state operator operators of the ancillary qubit with the value of the state operator on $\mathcal{H}$ attained over a completely positive evolution, the output is such that a measurement on the ancilla retrieves the initial value of the state operator on $\mathcal{H}$. We emphasize that this reversal protocol can be also applied to a completely positive evolution due to the action of a completely bounded flow on a special initial data, and therefore not described by a Lindblad-Gorini-Kossakowski-Sudarshan completely positive master equation. Reversal is exact for any initial data. This is at variance with protocols based on the Petz recovery map that can only reconstruct exactly a reference state but have the advantage of not requiring an ancilla \cite{KwMuKi2022}.
	
	In the theory of continuous time error correction \cite{PaZu1998,AhDoLa2002,AhWiMi2003,SaMi2005,HsBr2016} (see e.g. \cite{LiBr2013} for a general overview of quantum error correction), decoherence operators in the master equation model unwanted interactions with the environment of a quantum computing device. We thus conclude section~\ref{sec:recqc} by briefly discussing potential application of the recovery-by-embedding protocol to quantum error correction. 
	
	The application of the influence-martingale-induced embedding to the reversal of a quantum evolution is the third main contribution of this work.

	%Errors thus correspond to quantum jumps modeled by counting processes allowing only error channel per physical qubit at the time \cite{ShoP1995}. Perfectly detected errors induced by a weak continuous measurement can be thus corrected by quantum feedback protocols see e.g. \cite{AhWiMi2003}. These error correction models 
	
	%Our second application is to cancel out the effect of a reservoir on the unitary evolution of a system. 
	
	Finally, in section~\ref{sec:povm} we comment on the application of the influence martingale framework to certain non-canonical forms of the time-local master equation. We use section~\ref{sec:cb} and the appendices to fix the notation and recall some auxiliary results.

	\section{Canonical form of the time-local master equation} 
	\label{sec:cb}
	
	Our starting point is the discussion of the unique canonical form of time-local master equations in section II of \cite{HaCrLiAn2014}. 
	We summarize the  points relevant to the present work.
	We suppose that the open system is defined on a  Hilbert space $ \mathcal{H} $ of finite dimension $ d $.  
	The space $\mathcal{B}(\mathcal{H}) $ of bounded operators acting on $ \mathcal{H} $ then reduces to the set of  $ d\,\times\,d $ dimensional matrices $ \mathcal{M}_{d} $. We regard $ \mathcal{M}_{d} $ as a Hilbert space with respect to the Hilbert-Schimdt inner product \cite{BeZy2006}. 
	Any time-local master equation governing the evolution of the open system state operator $ \bm{\rho}_{t} $  is amenable to the form
	\begin{subequations}
		\label{cb:LGKS}
		\begin{align}
			&\label{cb:LGKS1}
			\partial_{t}\bm{\rho}_{t}=-\imath\,\left[\operatorname{H}_{t}\,,\bm{\rho}_{t}\right]
			+\sum_{\mathscr{l}=1}^{\mathscr{L}}\mathscr{w}_{\mathscr{l};t}\operatorname{D}_{\operatorname{L}_{\mathscr{l}; t}}(\bm{\rho}_{t})
			\\
			&
			\operatorname{D}_{\operatorname{L}_{\mathscr{l};t}}(\bm{\rho}_{t})=
						 \frac{1}{2}
			\left(
			\left[\operatorname{L}_{\mathscr{l}; t}\,,\bm{\rho}_{t}\operatorname{L}_{\mathscr{l}; t}^{\dagger}\right]
			+
			\left[\operatorname{L}_{\mathscr{l}; t}\bm{\rho}_{t}\,,\operatorname{L}_{\mathscr{l}; t}^{\dagger}\right]
			\right)
			\label{cb:LGKS2}
		\end{align}
	\end{subequations}
	The existence proof of (\ref{cb:LGKS}) has a long history see e.g. \cite{VstV1973,GrTaHa1977,vaWoLe1995}. It can be simply obtained from
	the Nakajima-Zwanzig equation (see e.g. \cite{ZwaR2001}) under the hypothesis that the solution map from an initial time $ \ti $ has a continuous inverse in an interval of finite duration \cite{AnCrHa2007,ChKo2010}. 
	The existence proof thus implies a parametric dependence of (\ref{cb:LGKS}) upon $ \ti $.  Multi-scale perturbation theory typically involve a coarse-grain of time scales \cite{WyBa1987}. Hence, in explicit derivations by time convolutionless perturbation theory \cite{ShTaHa1977,HaShSh1977,ChSh1979,BrPe2002,SmVa2010,KiWiRa2015,FeZiBl2021,NeBrWe2021} $ \ti $ can be approximated to zero. 
	
	In (\ref{cb:LGKS1}) we denote by $ \operatorname{H} $ the self-adjoint Hamiltonian operator, eventually time dependent. In the absence of the \textquotedblleft dissipator\textquotedblright\ $ \operatorname{D} $, the commutator in (\ref{cb:LGKS1}) is the generator of an unitary dynamics. The dissipator encapsulates non conservative interactions with the environment. In the canonical form (\ref{cb:LGKS2}), $ \operatorname{D} $ is  manifestly trace and self-adjointness preserving.  Distinguishing traits of the canonical form are \cite{HaCrLiAn2014}:
	\begin{enumerate}[nosep, leftmargin=10pt, label=\roman*]
		\item the sum ranges over $ \mathscr{L}=d^2-1 $ addends corresponding to a collection $ \left\{  \operatorname{L}_{\mathscr{l}; t}\right\}_{\mathscr{l}=1}^{\mathscr{L}} $ of decoherence, or Lindblad's, operators. At any time $ t $, the union of the collection together with the normalized  identity operator, constitutes an orthonormal basis of $ \mathcal{M}_{d} $ with respect to the Hilbert-Schimdt inner product:
		\begin{align}
			\operatorname{Tr} \operatorname{L}_{\mathscr{l}; t}=0\hspace{0.5cm}\&\hspace{0.5cm}
			\operatorname{Tr} \left(\operatorname{L}_{\mathscr{l}; t}^{\dagger}\operatorname{L}_{\mathscr{k}; t}\right)=\delta_{\mathscr{l},\mathscr{k}}
			\label{cb:ortho}
		\end{align} 
		\item the completeness relation in $ \mathcal{M}_{d} $ implies , that the decoherence operators satisfy the positive operator valued measurement type condition
		\begin{align}
			\sum_{\mathscr{l}=1}^{\mathscr{L}}\operatorname{L}_{\mathscr{l}; t}^{\dagger}\operatorname{L}_{\mathscr{l}; t}=\mathscr{g}\operatorname{1}_{\mathcal{H}}
			\label{cb:povm}
		\end{align}
		with $	\mathscr{g}=(d^{2}-1)/d $. We prove (\ref{cb:povm}) in appendix~\ref{ap:proof};
		\item the decoherence operators are unique modulo unitary transformations. The time dependence in general cannot be removed by unitary transformations unless $ d=2 $;
		\item \label{hyp:unique} the  time dependent functions $  \left\{ \mathscr{w}_{\mathscr{l};t} \right\}_{\mathscr{l}=1}^{\mathscr{L}}$ in (\ref{cb:LGKS2})
		are \emph{uniquely} determined by the canonical representation. They are referred to in \cite{HaCrLiAn2014} as canonical decoherence rates 
		a terminology that we adopt here too at variance with \cite{DoMG2022} where we called them \textquotedblleft weights\textquotedblright;
		\item  the canonical decoherence rates in (\ref{cb:LGKS2}) are \emph{not} sign definite. 
	\end{enumerate}
	In the present work we also surmise that canonical rates are sufficiently regular \emph{bounded functions}. 
	The latter assumption is expected to generically hold over finite time intervals \cite{vaWoLe2000}.
	From the mathematical point of view, (\ref{cb:LGKS}) is equivalent to a system of $ d^2-1 $ linear ordinary non-autonomous non-homogeneous differential equations as as can be verified by adopting the coherent vector representation \cite{HaCrLiAn2014}.
	Under the regularity assumptions we hypothesize, existence and uniqueness theorems of ordinary differential equations \cite{ArnV2006} ensure that we can construe any solution of (\ref{cb:LGKS}) as the application on the initial data of a two parameter linear flow 
	\begin{align}
		\mathscr{B}_{t,s}\colon \mathcal{M}_{d}\mapsto\mathcal{M}_{d}
		\nonumber
	\end{align}
	enjoying by definition in a time interval $ \mathbb{I} $ the group properties
	\begin{align}
		&\mathscr{B}_{t,s}= \mathscr{B}_{t,v}\mathscr{B}_{v,s} &\hspace{1.0cm}\forall\,t,v,s\in \mathbb{I}
		\nonumber\\
		&\mathscr{B}_{s,s}=\mathrm{Id}&\hspace{1.0cm}\forall\,s\in \mathbb{I}
		\nonumber
	\end{align}
	$ \mathrm{Id} $ being the identity map on $ \mathcal{M}_{d} $.
	The uniqueness of the canonical rates (property \ref{hyp:unique}) permits to rephrase the result of the axiomatic derivation of the Lindblad-Gorini-Kossakowski-Sudarshan master equation \cite{LinG1976,GoKoSu1976} by saying that the flow  is completely positive if and only if the canonical decoherence rates are positive definite. In such a case, the flow is always amenable to the canonical Choi-Stinespring form \cite{StiW1955,ChoM1975,KraK1983}
	\begin{align}
		\bm{\rho}_{t}=\mathscr{B}_{t,s}(\bm{\rho}_{s})=\sum_{a=1}^{\mathscr{N}}\operatorname{B}_{a; t,s}\bm{\rho}_{s}\operatorname{B}_{a; t,s}^{\dagger}
		\nonumber
	\end{align}
	for some $ \mathscr{N}\,\geq\,0 $ and a collection of two parameter families of operators $\operatorname{B}$ also satisfying
	\begin{align}
		\sum_{a=1}^{\mathscr{N}}\operatorname{B}_{a; t,s}^{\dagger}\operatorname{B}_{a; t,s}=\operatorname{1}_{\mathcal{H}}
		\nonumber
	\end{align}
	to enforce trace preservation. 
	
	Here, we only suppose that the canonical decoherence rates are bounded functions of time with at least one of them taking negative values over the time horizon of interest. In such a case, the flow is a completely bounded linear map \cite{PauV2003}. In the finite dimensional setup this means that completely bounded maps are the most general linear maps between matrix spaces \cite{JoKrPa2009}. The Wittstock-Paulsen decomposition \cite{WitG1981,PauV1982} yields the canonical form
	\begin{align}
		&\bm{\rho}_{t}=\mathscr{B}_{t,s}(\bm{\rho}_{s})
		\nonumber\\
		&=\sum_{a=1}^{\mathscr{N}^{(+)}}\operatorname{B}_{a; t,s}^{(+)}\bm{\rho}_{s}\operatorname{B}_{a; t,s}^{(+)\dagger}
		-\sum_{a=1}^{\mathscr{N}^{(-)}}\operatorname{B}_{a; t,s}^{(-)}\bm{\rho}_{s}\operatorname{B}_{a; t,s}^{(-)\dagger}
		\label{cb:WP}
	\end{align}
	as the difference of complete positive maps. In this case becomes
	\begin{align}
		\sum_{a=1}^{\mathscr{N}^{(+)}}\operatorname{B}_{a; t,\ti}^{(+)\dagger}\operatorname{B}_{a; t,\ti}^{(+)}
		=\operatorname{1}_{\mathcal{H}}+\sum_{a=1}^{\mathscr{N}^{(-)}}\operatorname{B}_{a; t,\ti}^{(-)\dagger}\operatorname{B}_{a; t,\ti}^{(-)}
		\nonumber
	\end{align}
	The Wittstock-Paulsen representation may also be interpreted as the diagonal form of the
	\textquotedblleft commutant\textquotedblright\ representation \cite{PaYuSu1985} (see also appendix~\ref{ap:commutant})
	\begin{align}
		\mathscr{B}_{t,s}(\bm{\rho}_{s})
		=\sum_{a,b=1}^{\tilde{\mathscr{N}}}C_{a,b; t,s}\operatorname{\tilde{B}}_{a; t,s}\bm{\rho}_{s}\tilde{\operatorname{B}}_{b; t,s}^{\dagger}
		\nonumber
	\end{align}
	The physical relevance of completely bounded master equations, i.e. canonical master equations with non-positive definite canonical rates stems from the following observations.  Completely positive evolution maps may be described as special solutions  of a completely bounded master equation. This is a situation encountered in models where the partial trace can be exactly evaluated  see e.g. \cite{TuZh2008,DoGoMG2020}. We recall the general mechanism in appendix~\ref{ap:flow}. More generally, complete positivity is only a sufficient condition \cite{JamA1972} that a linear operator map should satisfy in order to preserve positivity. In other words, positivity preserving maps over $ \mathcal{M}_{d} $, referred somewhat misleadingly as positive maps, do not need to have a positive spectrum when regarded as linear maps on a $ d^{2} $-dimensional vector space.  In particular, a positive map always admits the representation (\ref{cb:WP}) although the representation does not determine a-priori whether the map is indeed positive \cite{BeZy2006}.  It is, however, fair to add that from the foundational point of view, the interpretation of non-completely positive maps as dynamical maps is contentious see e.g. \cite{DoLi2016,ScRiSp2019} although their phenomenological use is widely accepted \cite{HaSt2020}.    
	
	We refer to \cite{SuSh2003,BeZy2006,PauV2003} for further properties of completely bounded maps.

	\section{Unraveling-paired complete positive map}
	\label{sec:unraveling}

	In \cite{DoMG2022} we prove that unraveling of a state operator solution of (\ref{cb:LGKS}) is amenable to the form
	\begin{align}
		\bm{\rho}_{t}=\operatorname{E}\left(\mu_{t}\bm{\psi}_{t}\bm{\psi}_{t}^{\dagger}\right)
		\label{unraveling:main}
	\end{align}
	where the expectation value is over a classical piecewise-deterministic process \cite{BrPe2002}. 	The logic underlying the proof of the unraveling is as follows.
	The interpretation of $ \bm{\psi}_{t} $ in (\ref{unraveling:main}) as a physical state vector requires it to take values on the Bloch hyper-sphere
	\begin{align}
		\left\|\bm{\psi}_{t}\right\|^{2}=1 
		\nonumber
	\end{align}
	with probability one.
	As a consequence, trace preservation imposes that $ \mu_{t} $ must be a mean value preserving martingale adapted to the natural filtration $ \left\{ \mathfrak{F}_{t} \right\}_{t\,\geq\,\ti} $ \cite{KleF2005} of 
	the process $ \left\{ \bm{\psi}_{t},\bm{\psi}_{t}^{\dagger}\right\}_{t\,\geq\,\ti} $. In other words, $ \mu_{t} $ is a functional of $ \bm{\psi}_{s}$, $\bm{\psi}_{s}^{\dagger} $ for any time $ s $ up to but no greater than $ t$. 
	As a consequence, at any instant of time it is always possible to define
	\begin{align}
		\mu_{t}^{(\pm)}=\max\left(0\,,\pm\mu_{t}\right)	
		\nonumber
	\end{align}
	in order to recover at statistical level a Wittstock-Paulsen decomposition (\ref{cb:WP})
	\begin{align}
		\bm{\rho}_{t}=\operatorname{E}\left(\mu_{t}^{(+)}\bm{\psi}_{t}\bm{\psi}_{t}^{\dagger}-\mu_{t}^{(-)}\bm{\psi}_{t}\bm{\psi}_{t}^{\dagger}\right)
		\nonumber
	\end{align}
	We now show that the proof of the unraveling holds true even under the more restrictive hypothesis than in \cite{DoMG2022} that the expectation value 
	\begin{align}
		\bm{\tilde{\rho}}_{t}=\operatorname{E}\bm{\psi}_{t}\bm{\psi}_{t}^{\dagger}
		\label{unraveling:dual}
	\end{align}
	is itself the solution of a Lindblad-Gorini-Kossakowski-Sudarshan master equation. We refer to (\ref{unraveling:dual}) as the completely positive \textquotedblleft unraveling-paired\textquotedblright\ state operator.
	We require $ \bm{\psi}_{t} $ to \emph{exactly} satisfy a non-linear, Bloch hyper-sphere preserving It\^o stochastic Schr\"odinger equation of the form \cite{BaBe1991,DaCaMo1992,BaHo1995}
	\begin{subequations}
		\label{unraveling:sse}
		\begin{align}
			&\label{unraveling:sse1}
			\mathrm{d}\bm{\psi}_{t}=\mathrm{d}t\,\bm{f}_{t}+\sum_{\mathscr{l}=1}^{\mathscr{L}}\mathrm{d}{\nu}_{\mathscr{l};t}\left(\frac{\operatorname{L}_{\mathscr{l}; t}\bm{\psi}_{t}}{\left\|\operatorname{L}_{\mathscr{l}; t}\bm{\psi}_{t}\right\|}-\bm{\psi}_{t}\right)
			\\
			&\label{unraveling:sse2}
			\bm{f}_{t}=-\imath\,\operatorname{H}_{t}\bm{\psi}_{t}-
			\sum_{\mathscr{l}=1}^{\mathscr{L}}\mathscr{r}_{\mathscr{l};t}\frac{\operatorname{L}_{\mathscr{l}; t}^{\dagger}\operatorname{L}_{\mathscr{l}; t}-\left\|\operatorname{L}_{\mathscr{l}; t}\bm{\psi}_{t}\right\|^{2}\operatorname{1}_{\mathcal{H}}}{2}\bm{\psi}_{t}
			\\
			&\label{unraveling:sse3}
			\bm{\psi}_{\ti}=\bm{z}
		\end{align}
	\end{subequations}
	with $ \bm{z}^{\dagger}\bm{z}=1 $ and 
	\begin{align}
		\mathscr{r}_{\mathscr{\mathscr{l}}; t} \,\geq\,0\hspace{0.5cm}\forall\,\mathscr{l}=1,\dots,\mathscr{L}
		\label{unraveling:rates}
	\end{align}
	positive definite rate functions whose explicit value will be determined below.
	The dynamics of $ \bm{\psi}_{t}^{\dagger} $ straightforwardly follows by applying the dual conjugation operation to (\ref{unraveling:sse}). The influence martingale process $ \left\{ \mu_{t} \right\}_{t\,\geq\,\ti} $ satisfies the It\^o stochastic differential equation 
	\begin{subequations}
		\label{martingale}
		\begin{align}
			&\label{martingale1}	\mathrm{d}\mu_{t}=\mu_{t}\sum_{\mathscr{l}=1}^{\mathscr{L}}\left(\frac{\mathscr{w}_{\mathscr{l};t}}{\mathscr{r}_{\mathscr{l};t}}-1\right)\,\mathrm{d}\iota_{\mathscr{l};t}
			\\
			&\label{martingale2}
			\mathrm{d}\iota_{\mathscr{l};t}=	\mathrm{d}\nu_{\mathscr{l};t}-\mathscr{r}_{\mathscr{l};t}\left\|\operatorname{L}_{\mathscr{l}; t}\bm{\psi}_{t}\right\|^{2}\mathrm{d}t
			\\
			&\label{martingale3}\mu_{\ti}=1
		\end{align}
	\end{subequations}
	In (\ref{unraveling:sse1}), (\ref{martingale2}) the $ \left\{\mathrm{d}{\nu}_{\mathscr{l};t}\right\}_{\mathscr{l}=1}^{\mathscr{L}} $ denote the increments of the independent unraveling processes \cite{BrPe2002,WiMi2009}:
	\begin{subequations}
		\label{unraveling:counting}
		\begin{align}
			&\label{unraveling:counting1}
			\mathrm{d}\nu_{\mathscr{l};t}\mathrm{d}\nu_{\mathscr{k},t}=\mathrm{d}\nu_{\mathscr{l};t}\mathrm{d}\iota_{\mathscr{k};t}=\delta_{\mathscr{l},\mathscr{k}}\mathrm{d}\nu_{\mathscr{l};t}
			\\
			&\label{unraveling:counting2}
			\operatorname{E}\big{(}\mathrm{d}\nu_{\mathscr{l};t}\big{|}\left\{ \bm{\psi}_{t}\,,\bm{\psi}_{t}^{\dagger} \right\}\cap\mathfrak{F}_{t}\big{)}
			=\mathscr{r}_{\mathscr{l};t}\left\|\operatorname{L}_{\mathscr{l}; t}\bm{\psi}_{t}\right\|^{2}\mathrm{d}t
		\end{align}
	\end{subequations}
	for $  \mathscr{l},\mathscr{k}=1,\dots,\mathscr{L}$. The stochastic differential in (\ref{martingale1}) are the compensated increments of the counting process (\ref{martingale1}). This fact together with boundedness assumptions on the rates $ \mathscr{r}_{\mathscr{l}; t} $ ensures that 
	$ \left\{ \mu_{t} \right\}_{r\,\geq\,\ti} $ enjoys the martingale property.
	
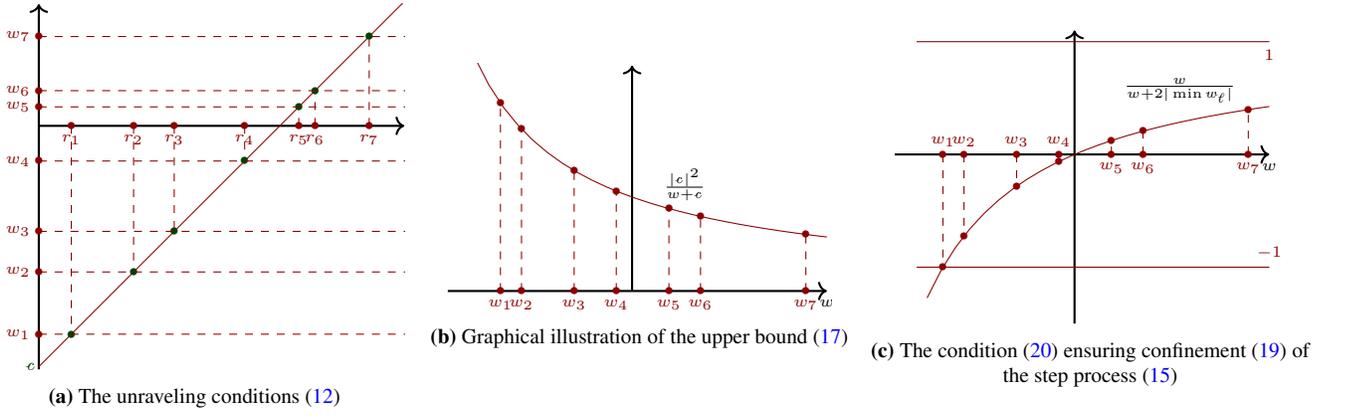
\begin{figure*}[tbh]
			\label{Fig:unraveling}
			\begin{subfigure}{0.33\textwidth}
				\centering
				\begin{tikzpicture}[xscale=0.36,yscale=0.36]
					\tkzInit[xmin=-2,xmax=13.5,ymin=-9,ymax=4.5]
					\tkzClip   
					\draw [thick, ->] (0,0) -- (13.5,0);
					\draw [thick, ->] (0,-9) -- (0,4.5);
					\draw [darkred,domain=0:14] plot (\x, {\x-8.9});
					\draw [dashed, thin, darkred] (8.9-7.7,0)node[below]{\textcolor{darkred}{$\scriptstyle{r}_{1}$}} node[]{\textcolor{darkred}{$\scriptstyle{\bullet}$}}--(8.9-7.7,-7.7)node[]{\textcolor{darkgreen}{$\scriptstyle{\bullet}$}};
					\draw [dashed, thin, darkred]  (8.9-5.4,0)node[below]{\textcolor{darkred}{$\scriptstyle{r}_{2}$}} node[]{\textcolor{darkred}{$\scriptstyle{\bullet}$}}--(8.9-5.4,-5.4)node[]{\textcolor{darkgreen}{$\scriptstyle{\bullet}$}};
					\draw [dashed, thin, darkred]  (8.9-3.9,0)node[below]{\textcolor{darkred}{$\scriptstyle{r}_{3}$}} node[]{\textcolor{darkred}{$\scriptstyle{\bullet}$}}--(8.9-3.9,-3.9)node[]{\textcolor{darkgreen}{$\scriptstyle{\bullet}$}};
					\draw [dashed, thin, darkred]  (8.9-1.3,0)node[below]{\textcolor{darkred}{$\scriptstyle{r}_{4}$}} node[]{\textcolor{darkred}{$\scriptstyle{\bullet}$}}--(8.9-1.3,-1.3)node[]{\textcolor{darkgreen}{$\scriptstyle{\bullet}$}};
					\draw [dashed, thin, darkred]  (8.9+0.7,0)node[below]{\textcolor{darkred}{$\scriptstyle{r}_{5}$}} node[]{\textcolor{darkred}{$\scriptstyle{\bullet}$}}--(8.9+0.7,0.7)node[]{\textcolor{darkgreen}{$\scriptstyle{\bullet}$}};
					\draw [dashed, thin, darkred]  (8.9+1.3,0)node[below]{\textcolor{darkred}{$\scriptstyle{r}_{6}$}} node[]{\textcolor{darkred}{$\scriptstyle{\bullet}$}}--(8.9+1.3,1.3)node[]{\textcolor{darkgreen}{$\scriptstyle{\bullet}$}};
					\draw [dashed, thin, darkred]  (8.9+3.3,0)node[below]{\textcolor{darkred}{$\scriptstyle{r}_{7}$}} node[]{\textcolor{darkred}{$\scriptstyle{\bullet}$}}--(8.9+3.3,3.3)node[]{\textcolor{darkgreen}{$\scriptstyle{\bullet}$}};
					\node at (-0.3,-8.9) {\textcolor{darkgreen}{$\scriptstyle{\mathscr{c}}$}};
					\draw [dashed, thin, darkred] (0,-7.7) node[left]{\textcolor{darkred}{$\scriptstyle{w}_{1}$}}node[]{\textcolor{darkred}{$\scriptstyle{\bullet}$}}--(14,-7.7);
					\draw [dashed, thin, darkred]  (0,-5.4) node[left]{\textcolor{darkred}{$\scriptstyle{w}_{2}$}}node[]{\textcolor{darkred}{$\scriptstyle{\bullet}$}}--(14,-5.4);
					\draw [dashed, thin, darkred]  (0,-3.9) node[left]{\textcolor{darkred}{$\scriptstyle{w}_{3}$}}node[]{\textcolor{darkred}{$\scriptstyle{\bullet}$}}--(14,-3.9);
					\draw [dashed, thin, darkred]  (0,-1.3) node[left]{\textcolor{darkred}{$\scriptstyle{w}_{4}$}}node[]{\textcolor{darkred}{$\scriptstyle{\bullet}$}}--(14,-1.3);
					\draw [dashed, thin, darkred]  (0,0.7) node[left]{\textcolor{darkred}{$\scriptstyle{w}_{5}$}}node[]{\textcolor{darkred}{$\scriptstyle{\bullet}$}}--(14,0.7);
					\draw [dashed, thin, darkred]  (0,1.3) node[left]{\textcolor{darkred}{$\scriptstyle{w}_{6}$}}node[]{\textcolor{darkred}{$\scriptstyle{\bullet}$}}--(14,1.3);
					\draw [dashed, thin, darkred]  (0,3.3) node[left]{\textcolor{darkred}{$\scriptstyle{w}_{7}$}}node[]{\textcolor{darkred}{$\scriptstyle{\bullet}$}}--(14,3.3);
				\end{tikzpicture}
				\caption{The unraveling conditions (\ref{unraveling:c})}
				\label{Fig:unraveling1}
			\end{subfigure}%
			\begin{subfigure}{0.33\textwidth}
				\centering
				\begin{tikzpicture}[xscale=0.7,yscale=0.25]
					\tkzInit[xmin=-3.5,xmax=3.78,ymin=-1,ymax=12.1]
					\tkzClip   
					\draw [thick, ->] (-3.5,0) -- (3.7,0)node[below]{$\scriptstyle{w}$};
					\draw [thick, ->] (0,0) -- (0,12);
					\draw [darkred,domain=-3:3.7] plot (\x, {25/(\x+5)});
					\node  at (1.0, {25/6+1.5}) {$\scriptstyle{\frac{|\mathscr{c}|^{2}}{w+\mathscr{c}}}$};
					\draw [dashed, thin, darkred] (-2.5,0) node[below]{\textcolor{darkred}{$\scriptstyle{w}_{1}$}}node[]{\textcolor{darkred}{$\scriptstyle{\bullet}$}}--(-2.5,{25/2.5})node[]{\textcolor{darkred}{$\scriptstyle{\bullet}$}};
					\draw [dashed, thin, darkred]  (-2.1,0) node[below]{\textcolor{darkred}{$\scriptstyle{w}_{2}$}}node[]{\textcolor{darkred}{$\scriptstyle{\bullet}$}}--(-2.1,{25/(-2.1+5)})node[]{\textcolor{darkred}{$\scriptstyle{\bullet}$}};
					\draw [dashed, thin, darkred]  (-1.1,0) node[below]{\textcolor{darkred}{$\scriptstyle{w}_{3}$}}node[]{\textcolor{darkred}{$\scriptstyle{\bullet}$}}--(-1.1,{25/(-1.1+5)})node[]{\textcolor{darkred}{$\scriptstyle{\bullet}$}};
					\draw [dashed, thin, darkred]  (-0.3,0) node[below]{\textcolor{darkred}{$\scriptstyle{w}_{4}$}}node[]{\textcolor{darkred}{$\scriptstyle{\bullet}$}}--(-0.3,{25/(-0.3+5)})node[]{\textcolor{darkred}{$\scriptstyle{\bullet}$}};
					\draw [dashed, thin, darkred]  (0.7,0) node[below]{\textcolor{darkred}{$\scriptstyle{w}_{5}$}}node[]{\textcolor{darkred}{$\scriptstyle{\bullet}$}}--(0.7,{25/(0.7+5)})node[]{\textcolor{darkred}{$\scriptstyle{\bullet}$}};
					\draw [dashed, thin, darkred]  (1.3,0) node[below]{\textcolor{darkred}{$\scriptstyle{w}_{6}$}}node[]{\textcolor{darkred}{$\scriptstyle{\bullet}$}}--(1.3,{25/(1.3+5)})node[]{\textcolor{darkred}{$\scriptstyle{\bullet}$}};
					\draw [dashed, thin, darkred]  (3.3,0) node[below]{\textcolor{darkred}{$\scriptstyle{w}_{7}$}}node[]{\textcolor{darkred}{$\scriptstyle{\bullet}$}}--(3.3,{25/(3.3+5)})node[]{\textcolor{darkred}{$\scriptstyle{\bullet}$}};
				\end{tikzpicture}
				\caption{Graphical illustration of the upper bound (\ref{variance:ub})}
				\label{Fig:unraveling2}
			\end{subfigure}
			%\nolinebreak
			\begin{subfigure}{0.33\textwidth}
				\centering
				\begin{tikzpicture}[xscale=0.7,yscale=1.5]
					\tkzInit[xmin=-3.4,xmax=4,ymin=-1.5,ymax=1.15]
					\tkzClip   
					\draw [thick, ->] (-3.5,0) -- (3.7,0)node[below]{$\scriptstyle{w}$};
					\draw [thick, ->] (0,-1.5) -- (0,1.1);
					\draw [darkred,domain=-2.8:3.7] plot (\x, {\x/(\x+5)});
					\node at (2, {2/7+0.1})[above]{$\scriptstyle{\frac{w}{w+2|\min \mathscr{w}_{\mathscr{l}}|}}$};
					\draw [darkred,domain=-3:3.7] plot (\x, {-1})node[above]{$\scriptstyle{-1}$};
					\draw [darkred,domain=-3:3.7] plot (\x, {1})node[below]{$\scriptstyle{1}$};
					\draw [dashed, thin, darkred] (-2.5,0) node[above]{\textcolor{darkred}{$\scriptstyle{w}_{1}$}}node[]{\textcolor{darkred}{$\scriptstyle{\bullet}$}}--(-2.5,{-2.5/2.5})node[]{\textcolor{darkred}{$\scriptstyle{\bullet}$}};
					\draw [dashed, thin, darkred]  (-2.1,0) node[above]{\textcolor{darkred}{$\scriptstyle{w}_{2}$}}node[]{\textcolor{darkred}{$\scriptstyle{\bullet}$}}--(-2.1,{-2.1/(-2.1+5)})node[]{\textcolor{darkred}{$\scriptstyle{\bullet}$}};
					\draw [dashed, thin, darkred]  (-1.1,0) node[above]{\textcolor{darkred}{$\scriptstyle{w}_{3}$}}node[]{\textcolor{darkred}{$\scriptstyle{\bullet}$}}--(-1.1,{-1.1/(-1.1+5)})node[]{\textcolor{darkred}{$\scriptstyle{\bullet}$}};
					\draw [dashed, thin, darkred]  (-0.3,0) node[above]{\textcolor{darkred}{$\scriptstyle{w}_{4}$}}node[]{\textcolor{darkred}{$\scriptstyle{\bullet}$}}--(-0.3,{-0.3/(-0.3+5)})node[]{\textcolor{darkred}{$\scriptstyle{\bullet}$}};
					\draw [dashed, thin, darkred]  (0.7,0) node[below]{\textcolor{darkred}{$\scriptstyle{w}_{5}$}}node[]{\textcolor{darkred}{$\scriptstyle{\bullet}$}}--(0.7,{0.7/(0.7+5)})node[]{\textcolor{darkred}{$\scriptstyle{\bullet}$}};
					\draw [dashed, thin, darkred]  (1.3,0) node[below]{\textcolor{darkred}{$\scriptstyle{w}_{6}$}}node[]{\textcolor{darkred}{$\scriptstyle{\bullet}$}}--(1.3,{1.3/(1.3+5)})node[]{\textcolor{darkred}{$\scriptstyle{\bullet}$}};
					\draw [dashed, thin, darkred]  (3.3,0) node[below]{\textcolor{darkred}{$\scriptstyle{w}_{7}$}}node[]{\textcolor{darkred}{$\scriptstyle{\bullet}$}}--(3.3,{3.3/(3.3+5)})node[]{\textcolor{darkred}{$\scriptstyle{\bullet}$}};
				\end{tikzpicture}
				\caption{The condition (\ref{variance:c}) ensuring confinement (\ref{variance:confined}) of the step process (\ref{variance:step})}
				\label{Fig:unraveling3}
			\end{subfigure}
			\centering
			\caption{Graphical proof of the unraveling and its consequences. }
		\end{figure*}

	According to these definitions, it is an immediate consequence of \cite{BaBe1991,DaCaMo1992,CarH1993,BrPe1995,WisH1996} that the Lindblad-Gorini-Kossakowski-Sudarshan master equation governing (\ref{unraveling:dual}) is
	\begin{align}
		\label{unraveling:LGKSaux}
		\partial_{t}\bm{\tilde{\rho}}_{t}=-\imath\,\left[\operatorname{H}_{t}\,,\bm{\tilde{\rho}}_{t}\right]
		+\sum_{\mathscr{l}=1}^{\mathscr{L}}\mathscr{r}_{\mathscr{l}; t}\operatorname{D}_{\operatorname{L}_{\mathscr{l}; t}}(\bm{\tilde{\rho}}_{t})
	\end{align}
	In order to ensure that (\ref{unraveling:main}) indeed satisfies (\ref{cb:LGKS}) we impose the unraveling conditions
	\begin{align}
		\mathscr{w}_{\mathscr{l}; t}=\mathscr{r}_{\mathscr{l}; t}-\mathscr{c}_{t}
		\label{unraveling:c}
	\end{align}
	The self-consistency of (\ref{unraveling:c}) hinges upon the requirement that the positive-definite function $ \mathscr{c}_{t} $ must fulfill  the inequality 
	\begin{align}
		\mathscr{c}_{t} \,>\, -\min_{\mathscr{l}=1,\dots,\mathscr{L}}\mathscr{w}_{\mathscr{l}; t}\,\equiv\,\left |\mathscr{w}_{\mathscr{l}_{\star};  t} \right |
		\label{unraveling:constraint}
	\end{align}
	Figure~\ref{Fig:unraveling1} yields a graphical proof of the self-consistence of the unraveling conditions.
	
	If we now insert (\ref{unraveling:c}) into the drift (\ref{unraveling:sse2}) and use (\ref{cb:povm}) on the Bloch hyper-sphere
	we recover the form of the drift \emph{hypothesized} in \cite{DoMG2022}
	\begin{align}
		\bm{f}_{t}=-\imath\,\operatorname{H}_{t}\bm{\psi}_{t}-
		\sum_{\mathscr{l}=1}^{\mathscr{L}}\mathscr{w}_{\mathscr{l};t}\frac{\operatorname{L}_{\mathscr{l}; t}^{\dagger}\operatorname{L}_{\mathscr{l}; t}-\left\|\operatorname{L}_{\mathscr{l}; t}\bm{\psi}_{t}\right\|^{2}\operatorname{1}_{\mathcal{H}}}{2}\bm{\psi}_{t}
		\label{unraveling:drift}
	\end{align}
	A straightforward application of It\^o lemma as done in \cite{DoMG2022} completes the proof. For readers' convenience we reproduce the steps of the calculation in appendix~\ref{ap:Ito}.  
	
	Some remarks are here in order.
	First, time dependence of  decoherence rates and operators does not play any role in the proof. Second, the unraveling (\ref{unraveling:sse}) 
	models a continuous time record of indirect measurements of the system gathered by means of decoherence channels consistent with the orthonormal conditions (\ref{cb:ortho}). We emphasize, however, that only the weaker condition (\ref{cb:povm}) is needed for (\ref{unraveling:main}) and (\ref{unraveling:dual}) to hold simultaneously true. Hence, the unraveling also holds when the time-local master (\ref{cb:LGKS}) is not in canonical form (i.e. $ \mathscr{L} $ is arbitrary and the conditions (\ref{cb:ortho}) are not satisfied) if (\ref{cb:povm}) is satisfied. In fact, in section~\ref{sec:povm} we show how to release (\ref{cb:povm}) in the non-canonical setup. The drawback of non-canonical master equations is that the signs of the rates do not immediately characterize the properties of the flow as linear operator map \cite{HaCrLiAn2014}.
	
	%   \pagebreak
	%
	%
	%	\onecolumngrid
	%
	%
	%	\begin{center}
		%	\end{center}
	%
	%
	%	\twocolumngrid
	%
	%
	
	\section{An optimization criterion for the influence martingale}
	\label{sec:variance}
	
	The It\^o stochastic differential equation governing the influence martingale is enslaved to the stochastic Schr\"odinger equation (\ref{embedding:sse}) and exactly integrable given the solution of this latter equation. In particular, for any physical path of $ \bm{\psi}_{t} $ on the Bloch hyper-sphere we get
	\begin{align}
		\mu_{t}=e^{\mathscr{g}\int_{\ti}^{t}\mathrm{d}s\,\mathscr{c}_{s}}\,\lambda_{t}
		\nonumber
	\end{align}
	where $ \lambda_{t} $ is the step process satisfying the It\^o stochastic differential equation
	\begin{subequations}
		\label{variance:step}
		\begin{align}
			&\label{variance:step1}
			\mathrm{d}\lambda_{t}=-\mathscr{c}_{t}\,\lambda_{t}\sum_{\mathscr{l}=1}^{\mathscr{L}}\frac{\mathrm{d}\nu_{\mathscr{l};t}}{\mathscr{w}_{\mathscr{l}; t}+\mathscr{c}_{t}}
			\\
			&\label{variance:step2}
			\lambda_{\ti}=1
		\end{align}
	\end{subequations}
	So far the the positive function $ \mathscr{c}_{t} $ in the unraveling conditions (\ref{unraveling:c}) is arbitrary and only subject to the constraint (\ref{unraveling:constraint}). The interpretation of the variance of the influence martingale suggests a criterion to resolve 
	such indeterminacy. Namely, let us consider
	\begin{align}
		\bm{\omega}_{t}=(\mu_{t}-1)\,\bm{\psi}_{t}\bm{\psi}_{t}^{\dagger}
		\nonumber
	\end{align}
	and its variance
	\begin{align}
		0\,\leq\,\operatorname{Tr}\operatorname{E}\left(\bm{\omega}_{t} -\operatorname{E}\bm{\omega}_{t}\right)^{2}
		=\operatorname{Tr}\operatorname{E}\bm{\omega}_{t} ^{2}-\operatorname{Tr}(\operatorname{E}\bm{\omega}_{t})^{2}
		\nonumber
	\end{align}
	We readily arrive at the inequality
	\begin{align}
		\operatorname{Tr}(\bm{\rho}_{t}-\bm{\tilde{\rho}}_{t})^{2}\,\leq\,\operatorname{E}\mu_{t}^{2}-1
		\label{variance:distance}
	\end{align}
	stating that the variance of the influence martingale yields an upper-bound on the squared Hilbert-Schmidt distance between the completely bounded and the unraveling-paired  completely positive state operator. From the information-theoretic point of view we may interpret the variance of the influence martingale as a $ \chi $-squared divergence \cite{TsyA2008} between bounded (pseudo-probability) measures. 
	
	The dynamics of the second moment of the influence martingale allows us to delve deeper on the consequences of the inequality (\ref{variance:distance}). It\^o lemma 
	yields
	\begin{align}
		\operatorname{E}\mathrm{d}\mu_{t}^{2}
		\,=\,\sum_{\mathscr{l}=1}^{\mathscr{L}}\frac{\mathscr{c}_{t}^{2}}{\mathscr{w}_{\mathscr{l}; t}+\mathscr{c}_{t}}(\operatorname{E}\mu_{t}^{2}\,\left\|\operatorname{L}_{\mathscr{l}; t}\bm{\psi}_{t}\right\|^{2})\mathrm{d}t
		\nonumber
	\end{align}
	By  (\ref{unraveling:constraint}) (see Fig.~\ref{Fig:unraveling2}) the right hand side is positive definite and admits an universal upper bound with respect to the state of the system. Namely, upon using \eqref{cb:povm} the sum reduces to
	\begin{align}
		\operatorname{E}\mathrm{d}\mu_{t}^{2}\,\leq\,\frac{\mathscr{g}\,\mathscr{c}_{t}^{2}\,\operatorname{E}\mu_{t}^{2}}{-|\mathscr{w}_{\mathscr{l}_{\star}; t}|+\mathscr{c}_{t}}\mathrm{d}t
		\label{variance:ub}
	\end{align}
	The choice
	\begin{align}
		\mathscr{c}_{t}^{\star} =2\,|\mathscr{w}_{\mathscr{l}_{\star}; t}|
		\label{variance:optimal}
	\end{align}
	yields the minimum upper bound in (\ref{variance:ub}) and as a consequence an analytic estimate of the variance of the influence martingale. Notably, for a completely positive master equation (\ref{variance:optimal}) is consistent with the reduction of the martingale to a dispersion-less
	process and thus the vanishing of the $ \chi $-squared divergence.
	
	A further consequence of (\ref{variance:optimal}) is the confinement of the step process (\ref{variance:step}):
	\begin{align}
		|\lambda_{t}|\,\leq\,1\hspace{0.5cm}\forall\,t\,\geq\,\ti
		\label{variance:confined}
	\end{align}
	To prove confinement we observe that a jump of  the $ \mathscr{l} $-counting processes on the right hand side of (\ref{variance:step1}) yields
	\begin{align}
		\lambda_{t+\mathrm{d}t}=\frac{\mathscr{w}_{\mathscr{l}; t}}{\mathscr{w}_{\mathscr{l}; t}+2\,|\mathscr{w}_{\mathscr{l}_{\star}; t}|}\lambda_{t}
		\nonumber
	\end{align}
	where by definition (see Fig~\ref{Fig:unraveling3})
	\begin{align}
		\left |\frac{\mathscr{w}_{\mathscr{l}; t}}{\mathscr{w}_{\mathscr{l}; t}+2\,|\mathscr{w}_{\mathscr{l}_{\star}; t}|}
		\right |\,\leq\,1
		\label{variance:c}
	\end{align}
	The confinement in the unit interval of the absolute value of the step process allows us to invoke the \emph{commutant representation} of a completely bounded map \cite{PaYuSu1985} (see also appendix~\ref{ap:commutant}) to conclude that the 
	evolution of the stochastic state operator in  $ \mathcal{M}_{2\,d} $
	\begin{align}
		\bm{\varsigma}_{t} =   \frac{\operatorname{1}_{2}+\lambda_{t}\,\sigma_{1}}{2}\,\otimes\,\bm{\psi}_{t}\bm{\psi}_{t}^{\dagger}
		\label{variance:extension}
	\end{align}
	is governed by a completely positive trace preserving map. In (\ref{variance:extension}) and below $ \sigma_{i} $, $ i=1,2,3 $ denote the Pauli matrices.
	
	\subsection{Relation with minimum rate of isotropic noise and  structural physical approximation}
	
	We can interpret the optimal criterion (\ref{variance:optimal}) as the condition that determines, under the constraints imposed by the unraveling, the closest	completely positive flow to that generated by a given completely bounded master equation. In this formulation, the optimization problem
	is reminiscent of the quantification of the distance of a snapshot of quantum evolution from a completely positive map studied in \cite{WoEiCuCi2008}.  In order to arrive to a computable quantifier, the authors of \cite{WoEiCuCi2008} introduce a version of the structural physical approximation \cite{HoEk2002} which leads to the definition of the minimum rate of isotropic noise $ \mathscr{n}_{t}^{\star} $. To exhibit the connection between these concepts, we refer to the derivation of the minimum rate of isotropic noise  presented in \cite{HaCrLiAn2014}.
	Over an infinitesimal time interval the flow of (\ref{cb:LGKS}) maps an arbitrary state operator $ \bm{\rho} $ as
	\begin{align}
		\mathscr{B}_{t+\mathrm{d}t,t}(\bm{\rho})=\bm{\rho}+\mathrm{d}t\,\mathcal{L}_{t}(\bm{\rho})
		\label{mrin:cb}
	\end{align}
	where $ \mathcal{L}_{t}(\bm{\rho}) $ denotes the right hand side of (\ref{cb:LGKS}). In general, (\ref{mrin:cb})  only describes a completely bounded evolution. To extricate from  (\ref{mrin:cb}) a completely positive map
	it is sufficient \cite{HoEk2002} to add completely depolarizing channel with a rate large enough to offset the most negative eigenvalue of the Choi matrix of (\ref{mrin:cb}). Explicitly, this means to determine the minimum value of $ \mathscr{n}_{t} $ such that
	\begin{align}
		&		\mathscr{P}_{t+\mathrm{d}t,t}[\mathscr{n}_{t}](\bm{\rho})=(1-\mathrm{d}t\,\mathscr{n}_{t})\mathscr{B}_{t+\mathrm{d}t,t}(\bm{\rho})+\mathrm{d}t\,\mathscr{n}_{t}\frac{\operatorname{1}_{\mathcal{H}}}{d}
		\nonumber\\
		&\approx \bm{\rho}+\mathrm{d}t\left(\mathcal{L}_{t}(\bm{\rho})+\mathscr{n}_{t}\left(\frac{\operatorname{1}_{\mathcal{H}}}{d}-\bm{\rho}\right)\right)
		\label{mrin:spa}
	\end{align}
	is completely positive. The analysis of the Choi matrix of (\ref{mrin:spa}) detailed in appendix~C of \cite{HaCrLiAn2014} proves that
	\begin{align}
		\mathscr{n}_{t}^{\star}=d\,|\mathscr{w}_{\mathscr{l}_{\star}; t}|=\frac{d}{2}\mathscr{c}_{t}^{\star}
		\nonumber
	\end{align}
	The optimization (\ref{mrin:spa}) corresponds to a structural physical approximation applied to the infinitesimal increment of a flow.
	We recall that in \cite{HoEk2002} Horodecki and Ekert introduce the structural physical approximation in connection with the positive map
	criterion stating that the state operator $ \bm{\rho} $ of a bipartite system is separable if and only if  
	\begin{align}
		(\mathrm{Id}\otimes \Lambda)(\bm{\rho})\,\geq\,0
		\nonumber
	\end{align}
	i.e. $ (\mathrm{Id}\otimes \Lambda)(\rho) $ is positive definite for all maps $ \Lambda $ positive but non-completely positive maps of operators
	acting on the Hilbert space of one of the constituents. The structural physical approximation corresponds to the minimal deformation of $ (\mathrm{Id}\otimes \Lambda) $ by a depolarizing map which occasion a completely positive maps. Hence, the structural physical approximation  opens the way to  laboratory realization of entaglement witness operations. We refer to \cite{ShuF2016} a recent review of this subject.

	\section{\textquotedblleft Embedding\textquotedblright\ quantum channel induced by the influence martingale}
	\label{sec:embedding}
	
	The question naturally arises whether the expectation value of (\ref{variance:extension}) specifies the solution of a Lindblad-Gorini-Kossakowski-Sudarshan master equation in $ \mathcal{M}_{2\,d} $.
	The answer is positive. To see this, we start by observing that the operator
	\begin{align}
		\bm{\varrho}_{t}=\operatorname{E}\lambda_{t}\bm{\rho}_{t}\bm{\rho}_{t}^{\dagger}=e^{-\mathscr{g}\int_{\ti}^{t}\mathrm{d}s\,\mathscr{c}_{s}}\bm{\rho}_{t}
		\nonumber
	\end{align}
	satisfies the non-trace preserving master equation
	\begin{align}
		&			\partial_{t}\bm{\varrho}_{t}=-\imath\,\left[\operatorname{H}_{t}\,,\bm{\varrho}_{t}\right]
		\nonumber\\
		&+	\sum_{\mathscr{l}=1}^{\mathscr{L}}\mathscr{r}_{\mathscr{l}; t}\operatorname{D}_{\operatorname{L}_{\mathscr{l}; t}}(\bm{\rho}_{t})-\mathscr{c}_{t}\sum_{\mathscr{l}=1}^{\mathscr{L}}\operatorname{L}_{\mathscr{l}; t}\bm{\varrho}_{t}\operatorname{L}_{\mathscr{l}; t}^{\dagger}
		\label{embedding:ntp}
	\end{align}
	implying
	\begin{align}
		|\operatorname{Tr}\bm{\varrho}_{t}|\,\leq\,1
		\nonumber
	\end{align}
	Next, we associate to each of the Lindblad operators in $ \mathcal{M}_{d}$ two operators in 
	$ \mathcal{M}_{2\,d}$ 
	\begin{align}
		&	\operatorname{V}_{2\mathscr{l}-1; t}=\operatorname{1}_{2}\,\otimes\,\operatorname{L}_{\mathscr{l}; t}	
		\nonumber\\
		&	\operatorname{V}_{2\mathscr{l}; t}= \sigma_{3}\,\otimes\,\operatorname{L}_{\mathscr{l}; t}			
		\nonumber
	\end{align}
	with corresponding rates
	\begin{align}
		& \mathscr{v}_{2\mathscr{l}-1; t}=\sqrt{\mathscr{r}_{\mathscr{l}; t}}\,
		\cos \frac{\theta_{\mathscr{l}; t}}{2} 
		\nonumber\\
		& \mathscr{v}_{2\mathscr{l}; t}=\sqrt{\mathscr{r}_{\mathscr{l}; t}}\,
		\sin \frac{\theta_{\mathscr{l}; t}}{2} 
		\nonumber
	\end{align}
	It is straightforward to verify that the operators $ \left\{ \operatorname{V}_{\mathscr{l}; t} \right\}_{\mathscr{l}=1}^{2\,\mathscr{L}} $satisfy the canonical relations (\ref{cb:ortho}) whereas the rates are positive for
	\begin{align}
		0\,\leq\,\theta_{\mathscr{l}; t}\,\leq\,\pi
		\label{embedding:cp}
	\end{align}
	As last step, we introduce the Lindblad-Gorini-Kossakowski-Sudarshan master equation for the state operator $\bm{\gamma}_{t}$ on $\mathbb{C}^2\otimes \mathcal{H}$
	\begin{align}
		&	\partial_{t}\bm{\gamma}_{t}=-\imath\left[\operatorname{1}_{2}\,\otimes\,\operatorname{H}_{t}\,,\bm{\gamma}_{t}\right]
+\sum_{\mathscr{l}=1}^{2\,\mathscr{L}}\mathscr{v}_{\mathscr{l}; t}
		\operatorname{D}_{\operatorname{V}_{\mathscr{l}; t}}\left(\bm{\gamma}_{t}\right)
%		\frac{\left[\operatorname{V}_{\mathscr{l}; t}\,,\bm{\gamma}_{t}\operatorname{V}_{\mathscr{l}; t}^{\dagger}\right]+\left[\operatorname{V}_{\mathscr{l}; t}\bm{\gamma}_{t}\,,\operatorname{V}_{\mathscr{l}; t}^{\dagger}\right]}{2}
		\label{embedding:LGKS}
	\end{align}
	Our definitions imply that
	\begin{align}
		\sum_{\mathscr{i}=1}^{2}\mathscr{v}_{2\mathscr{l}-2+\mathscr{i}; t}\operatorname{V}_{2\mathscr{l}-2+\mathscr{i}; t}^{\dagger}\operatorname{V}_{2\mathscr{l}-2+i; t}=
		\mathscr{r}_{\mathscr{l}; t} \operatorname{1}_{2}\,\otimes\,\operatorname{L}_{\mathscr{l}; t}^{\dagger}	\operatorname{L}_{\mathscr{l}; t}	
		\nonumber
	\end{align}
	and
	\begin{align}
		&	\sum_{\mathscr{i}=1}^{2}\mathscr{v}_{2\mathscr{l}-2+\mathscr{i}; t}\operatorname{V}_{2\mathscr{l}-2+\mathscr{i}; t}
		\begin{bmatrix}
			\bm{\gamma}_{1,1; t}& \bm{\gamma}_{1,2; t}\\   \bm{\gamma}_{2,1; t} & \bm{\gamma}_{2,2; t}
		\end{bmatrix}
		\operatorname{V}_{2\mathscr{l}-2+\mathscr{i}; t}^{\dagger}
		\nonumber\\
		&
		=\mathscr{r}_{\mathscr{l}; t} \begin{bmatrix}
			\operatorname{L}_{\mathscr{l}; t}\bm{\gamma}_{1,1; t}\operatorname{L}_{\mathscr{l}; t}^{\dagger}	& \cos \theta_{\mathscr{l}; t} \operatorname{L}_{\mathscr{l}; t}\bm{\gamma}_{1,2; t}\operatorname{L}_{\mathscr{l}; t}^{\dagger}\\   \cos \theta_{\mathscr{l}; t}\operatorname{L}_{\mathscr{l}; t}\bm{\gamma}_{2,1; t}\operatorname{L}_{\mathscr{l}; t}^{\dagger} & \operatorname{L}_{\mathscr{l}; t}\bm{\gamma}_{2,2; t}\operatorname{L}_{\mathscr{l}; t}^{\dagger}
		\end{bmatrix}
		\nonumber
	\end{align}
	Hence, we are always entitled to relate diagonal blocks of $ \bm{\gamma}_{t} $ to the solution of (\ref{unraveling:LGKSaux})
	e.g. by setting
	\begin{align}
		\bm{\gamma}_{1,1; t}=\bm{\gamma}_{2,2; t}=\frac{1}{2}\bm{\tilde{\rho}}_{t}
		\nonumber
	\end{align}
	Furthermore, the off-diagonal blocks satisfy (\ref{embedding:ntp}) 
	\begin{align}
		\bm{\gamma}_{1,2; t}=\bm{\gamma}_{2,1; t}=\frac{1}{2}\bm{\varrho}_{t}
		\nonumber
	\end{align}
	if the compatibility conditions
	\begin{align}
		\cos\theta_{\mathscr{l}; t}=\frac{\mathscr{r}_{\mathscr{l}; t}-\mathscr{c}_{t}}{\mathscr{r}_{\mathscr{l}; t}}=\frac{\mathscr{w}_{\mathscr{l}; t}}{\mathscr{w}_{\mathscr{l}; t}+\mathscr{c}_{t}}
		\label{embedding:cc}
	\end{align}
	hold true. This is the case if $ \mathscr{c}_{t} $ is equal to the \textquotedblleft optimal\textquotedblright\ value (\ref{variance:optimal})
	which implies the confinement condition (\ref{variance:c}). Furthermore the compatibility conditions (\ref{embedding:cc}) 
	admit unique solutions in the \textquotedblleft principal branch\textquotedblright\ specified by the requirement (\ref{embedding:cp}) of positive definite canonical rates.
	
\begin{figure}
		\centering
		\includegraphics[scale=0.8]{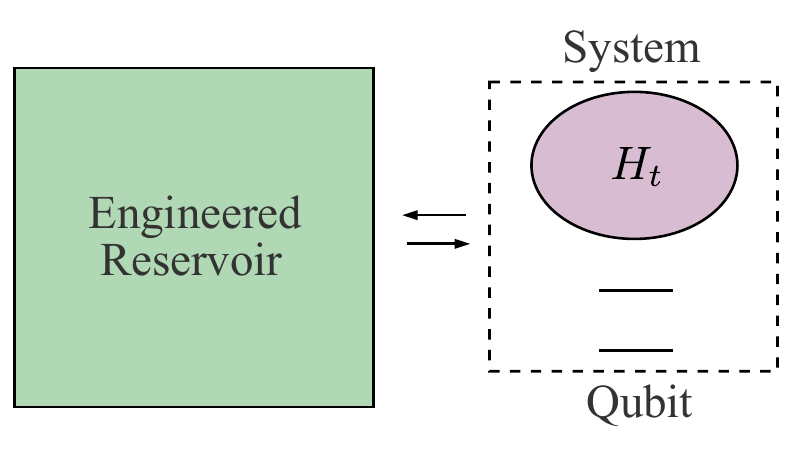}
		\caption{The master equation in Eq. \eqref{embedding:LGKS} represents the dynamics of the system and an auxiliary qubit. The system and qubit do not interact via a Hamiltonian and the qubit has no internal dynamics. However they do interact via the Lindblad operators $\operatorname{V}_{\ell;t}$ that could be realized by a specifically engineered reservoir.}
		\label{fig:embedding}
	\end{figure}

	To summarize, we have proven the identity
	\begin{align}
		\bm{\gamma}_{t}\,\equiv\,\frac{1}{2}\begin{bmatrix}
			\bm{\tilde{\rho}}_{t}	& \bm{\varrho}_{t}\\   \bm{\varrho}_{t} & \bm{\tilde{\rho}}_{t}
		\end{bmatrix}=\operatorname{E}\left(\frac{\operatorname{1}_{2}+\lambda_{t}\,\sigma_{1}}{2}\,\otimes\,\bm{\psi}_{t}\bm{\psi}_{t}^{\dagger}\right)
		\label{embedding:matrix}
	\end{align}
	stating that the solution of a completely bounded master equation and the solution of the influence-martingale-optimally-paired Lindblad-Gorini-Kossakowski-Sudarshan master can be always construed as generated by completely positive divisible map evolving state operators on an embedding Hilbert space $ \mathbb{C}^{2}\,\otimes\,\mathcal{H} $. Physically, (\ref{embedding:matrix}) describes the interaction of the original system with an ancillary qubit stylized in Fig~\ref{fig:embedding}. In this latter Hilbert space, by the known theory e.g. \cite{BaBe1991,DaCaMo1992} there also exists an unraveling of the form
	\begin{align}
		\bm{\gamma}_{t}=\operatorname{E}\bm{\Psi}_{t}\bm{\Psi}_{t}^{\dagger}
		\label{embedding:unraveling}
	\end{align}
	with $ \bm{\Psi} _{t}$ a state vector solution of the It\^o stochastic differential equation
	\begin{subequations}
		\label{embedding:sse}
		\begin{align}
			&\label{embedding:sse1}
			\mathrm{d}\bm{\Psi}_{t}=\mathrm{d}t\,\bm{F}_{t}+\sum_{\mathscr{l}=1}^{2\mathscr{L}}\mathrm{d}{N}_{\mathscr{l};t}\left(\frac{\operatorname{V}_{\mathscr{l}; t}\bm{\Psi}_{t}}{\left\|\operatorname{V}_{\mathscr{l}; t}\bm{\Psi}_{t}\right\|}-\bm{\Psi}_{t}\right)
			\\
			&\label{embedding:sse2}
			\bm{F}_{t}=-\imath\,\operatorname{1}_{2}\,\otimes\,\operatorname{H}_{t}\bm{\Psi}_{t}
			\nonumber\\
			&		-
			\sum_{\mathscr{l}=1}^{2\mathscr{L}}\mathscr{v}_{\mathscr{l}; t}\frac{\operatorname{V}_{\mathscr{l}; t}^{\dagger}\operatorname{V}_{\mathscr{l}; t}-\left\|\operatorname{V}_{\mathscr{l}; t}\bm{\Psi}_{t}\right\|^{2}\operatorname{1}_{\mathbb{C}^{2}\otimes \mathcal{H}}}{2}\bm{\Psi}_{t}
		\end{align}
	\end{subequations}
	driven by counting processes now characterized by
	\begin{align}
		&	\mathrm{d}N_{\mathscr{l};t}\mathrm{d}N_{\mathscr{k},t}=\delta_{\mathscr{l},\mathscr{k}}\mathrm{d}N_{\mathscr{l};t}
		\nonumber\\
		&	\operatorname{E}\left(\mathrm{d}N_{\mathscr{l};t}\big{|}\left\{ \bm{\Psi}_{t}\,,\bm{\Psi}_{t}^{\dagger} \right\}\cap\tilde{\mathfrak{F}}_{t}\right)=\mathscr{v}_{\mathscr{l}; t}\left\|\operatorname{V}_{\mathscr{l}; t}\bm{\Psi}_{t}\right\|^{2}\mathrm{d}t
		\nonumber
	\end{align}
	$ \left\{  \tilde{\mathfrak{F}}_{s} \right\}_{s\,\geq\,\ti}$ being the natural filtration of $ \left\{  \bm{\Psi}_{s}\,, \bm{\Psi}_{s}^{\dagger} \right\}_{s\,\geq\,\ti} $.
	
	It is worth emphasizing that \cite{BrKaPe1999,BreH2004} also introduce an embedding Hilbert space, and specifically $ \mathbb{C}^{3}\,\otimes\,\mathcal{H} $, to unravel by means of (\ref{embedding:sse}) the solution $ \bm{\rho}_{t} $ of a completely bounded master equation. The influence martingale has the double advantage of requiring a smaller dimension of the Hilbert space and to give direct access to off diagonal blocks thus overcoming the need to introduce (\ref{embedding:sse}). 
	
	One motivation to the introduce the embedding representation is to provide an avenue for a continuous time measurement interpretation \cite{BaLu2005} even in the case when $ \bm{\rho}_{t} $ is non positive \cite{BreH2004, DoMG2022}. In the coming section, we analyze the application to time reversal of a completely positive evolution.

	\subsection{Remark on the Paulsen-Suen \textquotedblleft triple\textquotedblright norm and optimal definition of embedding}
	\label{sec:norm}
	
	In \cite{PaYuSu1985} the examination of order theoretic characterizations of the norm for completely bounded maps leads Paulsen-Suen to
	introduce a new norm, referred to as \textquotedblleft triple\textquotedblright\ in \cite{PauV2007}, for a completely bounded map $ \mathscr{B} $. The definition the norm relies on the idea of finding an optimal auxiliary completely positive map and trace preserving $\mathscr{P} $ permitting to embed $ \mathscr{B} $
	in a completely positive map $ \mathscr{E} $ defined by
	\begin{align}
		\mathscr{E}\left(\begin{bmatrix}
			\operatorname{Y}	& \operatorname{X} \\\operatorname{W}	& \operatorname{Z} 
		\end{bmatrix}\right)
		=
		\begin{bmatrix}
			\mathscr{P}(\operatorname{Y})	& \mathscr{B}(\operatorname{X}) \\ \mathscr{B}^{\dagger}(\operatorname{W}	)& \mathscr{P}(\operatorname{Z} )
		\end{bmatrix}
		\nonumber
	\end{align}
	Paulsen and Suen triple norm is thus
	\begin{align}
		||| \mathscr{B} |||=\inf \left\{ \left\|\mathscr{P}\right\| \,\mbox{such that}\,\,\mathscr{E} \,\mbox{is completely positive}\right\}
		\label{norm:dominating}
	\end{align}
	In the definition the infimum is over the operator  norm $ \left\|\cdot\right\|_{\mathrm{cb}} $. In the finite dimensional case, the use of the canonical operator sum representation of $ \mathscr{P}\colon \mathcal{M}_{d}\mapsto \mathcal{M}_{d}$ for some  integer $ \mathscr{m} $
	\begin{align}
		\mathscr{P}(\operatorname{X})=\sum_{\mathscr{i}=1}^{\mathscr{m}}\operatorname{A}_{i}\operatorname{X}\operatorname{A}_{i}^{\dagger}
		\nonumber
	\end{align}
	yields
	\begin{align}
		\left\|\mathscr{P}\right\|:=\sup_{\bm{v}\in\,\mathcal{H}\,|\,\left\|v\right\|\,\leq\,1}
		\left\|\sum_{\mathscr{i}=1}^{\mathscr{m}}\operatorname{A}_{\mathscr{i}}\operatorname{A}_{\mathscr{i}}^{\dagger}\bm{v}\right\|
		\nonumber
	\end{align}
	A completely positive map satisfying $ ||| \mathscr{B} |||= \left\|\mathscr{P}\right\|$ is called \textquotedblleft dominating\textquotedblright\ \cite{PaYuSu1985}. 
	
	It has not escaped our notice that the optimization of the variance of the influence martingale yields a value (\ref{variance:optimal}) that is threshold to satisfy the embedding conditions (\ref{embedding:cc}). Larger values of $\mathscr{c}_{t}$ would be consistent with the embedding at the price of larger values of the rates $\mathscr{r}_{\mathscr{l}; t}$ and consequently of the operator norm of the realizations of the completely positive map in the diagonal blocks.
	The question naturally arises whether restricting the optimization in the definition of the triple norm to flows of master equations is equivalent to the optimization we consider.
	We leave proving or disproving this conjecture to future work.
	
	\section{Recovery of an initial state operator}
	\label{sec:recovery}
	
	The influence martingale framework unravels any completely bounded master equations.
	An appealing application is time reversal of a completely positive evolution in a finite time horizon $ \left[\ti\,,\tf\right] $. 
	We start from the completely positive canonical master equation  ($ \mathscr{h}_{\mathscr{l},t}\,\geq\,0 $)
	\begin{subequations}
		\label{recovery:fme}
		\begin{align}
			&\label{recovery:fme1}
			\bm{\dot{\chi}}_{t}=-\imath\,\left[\operatorname{H}_{t}\,,\bm{\chi}_{t}\right]+\sum_{\mathscr{l}=1}^{\mathscr{L}}\mathscr{h}_{\mathscr{l},t}\operatorname{D}_{ \operatorname{L}_{\mathscr{l}; t} }(\bm{\chi}_{t})
			\\
			&\label{recovery:fme2}
			\bm{\chi}_{\ti}=\bm{\chi}_{\iota}
		\end{align}
	\end{subequations}
	and consider the involution of the time parameter
	\begin{align}
		t^{\flat}=\tf+\ti-t
		\label{recovery:involution}
	\end{align}
	There are now two avenues to describe the evolution of the inverse of the flow of (\ref{recovery:fme})
	
	\subsection{Genuine backward dynamics}
	
	We define the reverse of the solution of (\ref{recovery:fme}) as
	\begin{align}
		\bm{\chi}_{t^{\flat}}^{\flat}=\bm{\chi}_{t}
		\nonumber
	\end{align}
	We use the \textquotedblleft overdot\textquotedblright\ notation to indicate differentiation with respect to the explicit time dependence:
	\begin{align}
		\bm{\dot{\chi}}_{t^{\flat}}^{\flat}\,\equiv\,\frac{\mathrm{d}}{\mathrm{d}t^{\flat} }\bm{\chi}_{t^{\flat}}^{\flat}=\frac{\mathrm{d}t}{\mathrm{d} t^{\flat} }\bm{\dot{\chi}}_{t}=-\bm{\dot{\chi}}_{t}
		\nonumber
	\end{align}
	We thus obtain the backward master equation
	\begin{subequations}
		\label{recovery:bme}
		\begin{align}
			&\label{recovery:bme1}
			\bm{\dot{\chi}}_{t^{\flat}}^{\flat}=\imath\,\left[\operatorname{H}_{t^{\flat}}^{\flat}\,,\bm{\chi}_{t^{\flat}}^{\flat}\right]-
			\sum_{\mathscr{l}=1}^{\mathscr{L}}\mathscr{h}_{\mathscr{l}; t^{\flat}}^{\flat}\operatorname{D}_{ \operatorname{L}_{\mathscr{l}; t^{\flat}} }(\bm{\chi}_{t^{\flat}}^{\flat})
			\\
			&\label{recovery:bme2}
			\bm{\chi}_{\tf}^{\flat}=\bm{\chi}_{\tf}
		\end{align}
	\end{subequations}
	where
	\begin{align}
		&\operatorname{H}_{t^{\flat}}^{\flat}=\operatorname{H}_{t}
		\nonumber\\
		&\mathscr{h}_{\mathscr{l};t^{\flat}}^{\flat}=\mathscr{h}_{\mathscr{l}; t}
		\nonumber
	\end{align}
	To unravel (\ref{recovery:bme}) we introduce the descending filtration $ \left\{  \mathfrak{F}_{s}^{\flat}\right\}_{s\,\leq\,\tf} $of \textquotedblleft future events\textquotedblright\ \cite{ChZa2003} i.e. a sequence of $ \chi $-algebras increasing as the time $ t^{\flat} $ \emph{decreases} from $ \tf $ to $ \ti $. Correspondingly, we consider the backward stochastic differential equation \cite{NelE2001} in the \emph{post-point} prescription
	\begin{align}
		&	\mathrm{d}^{\flat}\bm{\psi}_{t^{\flat}}^{\flat}=\bm{\psi}_{t^{\flat}}^{\flat}-\bm{\psi}_{t^{\flat}-\mathrm{d}t}^{\flat}
		\nonumber\\
		&=	\left(\imath \,\operatorname{H}_{t^{\flat}}^{\flat}-
		\sum_{\mathscr{l}=1}^{\mathscr{L}} \mathscr{h}_{\mathscr{l};t^{\flat}}^{\flat} \frac{\operatorname{L}_{\mathscr{l}; t}^{\dagger}\operatorname{L}_{\mathscr{l}; t}-\left\|\operatorname{L}_{\mathscr{l}; t}\bm{\psi}_{t^{\flat}}^{\flat}\right\|^{2}}{2}\right)\bm{\psi}_{t^{\flat}}^{\flat}\mathrm{d}t
		\nonumber\\
		&	-
		\sum_{\mathscr{l}=1}^{\mathscr{L}}\mathrm{d}\nu_{\mathscr{l}; t^{\flat}}^{\flat}\left(\frac{\operatorname{L}_{\mathscr{l}; t}\bm{\psi}_{t^{\flat }}^{\flat}}{\left\|\operatorname{L}_{\mathscr{l}; t}\bm{\psi}_{t^{\flat }}^{\flat}\right\|}
		-\bm{\psi}_{t^{\flat }}^{\flat}\right)
		\label{recovery:bsde}
	\end{align}
	The differentials
	\begin{align}
		\mathrm{d}\nu_{\mathscr{l}; t^{\flat}}^{\flat}=\nu_{\mathscr{l}; t^{\flat}}^{\flat}-\nu_{\mathscr{l}; t^{\flat}-\mathrm{d}t}^{\flat}
		\nonumber
	\end{align}
	of the counting processes in (\ref{recovery:bsde}) satisfy the same differential algebra relations of forward increments (\ref{unraveling:counting1}). 
	In consequence of the post-point prescription we require that counting process differentials are characterized by conditional expectations with respect to the descending filtration
	\begin{align}
		\operatorname{E}\left(\mathrm{d}^{\flat}\nu_{\mathscr{l}; t^{\flat}}^{\flat}\big{|}\left\{ \bm{\psi}_{t^{\flat }}^{\flat}\,,\bm{\psi}_{t^{\flat }}^{\flat \dagger} \right\}\cap\mathfrak{F}_{t^{\flat }}^{\flat}\right)=\mathscr{h}_{t^{\flat } }^{\flat}\left\|\operatorname{L}_{\mathscr{l}; t}\bm{\psi}_{t^{\flat }}^{\flat}\right\|^{2}\mathrm{d}t
		\nonumber
	\end{align}
	Backwards It\^o differential formula 
	\begin{align}
		&	\mathrm{d}\bm{\chi}_{t^{\flat}}^{\flat}=\operatorname{\operatorname{E}}\mathrm{d}^{\flat}\left(\bm{\psi}_{t^{\flat}}^{\flat}\bm{\psi}_{t^{\flat}}^{\flat \dagger}\right)
		=
		\operatorname{E}\left(\mathrm{d}^{\flat} \bm{\psi}_{t^{\flat}}\right)^{\flat}\bm{\psi}_{t^{\flat}}^{\flat \dagger}
		\nonumber\\	
		&+
		\operatorname{E}\left( \bm{\psi}_{t^{\flat}}^{\flat}\mathrm{d}^{\flat}\bm{\psi}_{t^{\flat}}^{\flat\dagger}\right)
		-
		\operatorname{E}\left (\left(\mathrm{d}^{\flat} \bm{\psi}_{t^{\flat}}\right)\mathrm{d}^{\flat}\bm{\psi}_{t^{\flat}}^{\flat\dagger}\right)
		\nonumber
	\end{align}
	recovers (\ref{recovery:bme1}). The terminal condition (\ref{recovery:bme1}) is implemented by assigning the terminal condition on the state vector from a probability distribution such that
	\begin{align}
		\bm{\chi}_{\tf}=\operatorname{E}\bm{\psi}_{\tf}^{\flat}\bm{\psi}_{\tf}^{\flat \dagger}
		\nonumber
	\end{align}
	Clearly paths solution of (\ref{recovery:bsde}) are not reversed path of the unraveling of (\ref{recovery:fme}) but realization of a distinct stochastic process whose connection to (\ref{recovery:fme}) resides in the second order statistics:
	\begin{align}
		\bm{\chi}_{t}=\operatorname{E}\bm{\psi}_{t^{\flat}}^{\flat}\bm{\psi}_{t^{\flat}}^{\flat \dagger}
		\nonumber
	\end{align} 
	Although mathematically straightforward, from the physical point of view the existence of the process (\ref{recovery:bsde})  appears to be mostly of conceptual interest.
	
	\subsection{Forward implementation of the backward dynamics}
	\label{sec:fbdyn}
	
	In order to describe the reversed dynamics in terms of a time variable $ t $ increasing from $ \ti $ to $ \tf $
	we posit
	\begin{align}
		\bm{\chi}_{t}^{\flat}=\bm{\chi}_{\tf+\ti-t}
		\label{recovery:fbop}
	\end{align}
	In such a case, we arrive at the completely bounded canonical master equation
	\begin{subequations}
		\label{recovery:fbme}
		\begin{align}
			&\label{recovery:fbme1}
			\bm{\dot{\chi}}_{t}^{\flat}=\imath\,\left[\operatorname{H}_{\tf+\ti-t}\,,\bm{\chi}_{t}^{\flat}\right]-
			\sum_{\mathscr{l}=1}^{\mathscr{L}}\mathscr{h}_{\mathscr{l};\tf+\ti-t } \operatorname{D}_{\operatorname{L}_{\mathscr{l}; t}}(\bm{\chi}_{t}^{\flat})
			\\
			&\label{recovery:fbme2}
			\bm{\chi}_{\ti}^{\flat}=\bm{\chi}_{\tf}
		\end{align}	
	\end{subequations}
	The use of the influence martingale yields the identity
	\begin{align}
		\bm{\dot{\chi}}_{t}^{\flat}=e^{\mathscr{g}\int_{\ti}^{\tf}\mathrm{d}s\,\mathscr{c}_{s}}\operatorname{E}\lambda_{t}\bm{\psi}_{t}\bm{\psi}_{t}^{\dagger}
		\label{recovery:funravel}
	\end{align}
	where the state vector solves the forward It\^o stochastic differential equation (\ref{unraveling:sse}) once we perform in (\ref{unraveling:sse2}) 
	the replacement
	\begin{align}
		\operatorname{H}_{t}\to -\operatorname{H}_{\tf+\ti-t},
		\nonumber
	\end{align}
	we impose the unraveling conditions
	\begin{align}
		\mathscr{r}_{\mathscr{l};t}=\mathscr{c}_{t}-\mathscr{h}_{\mathscr{l};\tf+\ti-t}
		\nonumber
	\end{align}
	with
	\begin{align}
		\mathscr{c}_{t}=2\max_{\mathscr{l}=1,\dots,\mathscr{L}} \mathscr{h}_{\mathscr{l};\tf+\ti-t}
		\nonumber
	\end{align}
	and we sample the initial condition (\ref{unraveling:sse3}) from a probability such that (\ref{recovery:fbme2}) holds. 
	These requirements entail that the process $ \left\{ \lambda_{t} \right\}_{t\,\geq\,\ti} $ in (\ref{recovery:funravel}) satisfies the equation
	\begin{align}
		\mathrm{d}\lambda_{t}=-\lambda_{t}\sum_{\mathscr{l}=1}^{\mathscr{L}}\left(\frac{\mathscr{h}_{\mathscr{l};\tf+\ti-t}}{\mathscr{r}_{\mathscr{l};t}}+1\right)\mathrm{d}\nu_{\mathscr{l};t}
		\nonumber
	\end{align}
	subject as usual to the initial condition $ \lambda_{\ti}=1 $.
	
	We emphasize that the jumping rates of the forward dynamics unraveling (\ref{recovery:fbme}) in general differ from those unraveling the forward dynamics (\ref{recovery:fme}) we are reversing:
	\begin{align}
		\mathscr{r}_{\mathscr{l}; t} \neq \mathscr{h}_{\mathscr{l}; t}
		\nonumber
	\end{align}
	If however, the microscopic dynamics yields
	\begin{align}
		\mathscr{h}_{\mathscr{l}; t}=1\hspace{0.5cm}\forall\,\mathscr{l}=1,\dots,\mathscr{L}\hspace{0.2cm}\&\hspace{0.2cm}t\,\in[\ti\,,\tf]
		\nonumber
	\end{align}
	we get immediately
	\begin{align}
		\mathscr{r}_{\mathscr{l}; t}=1\hspace{0.5cm}\forall\,\mathscr{l}=1,\dots,\mathscr{L}\hspace{0.2cm}\&\hspace{0.2cm}t\,\in[\ti\,,\tf]
		\nonumber
	\end{align} 
	This situation maybe encountered for an open system dynamics brought about by an environment described by an equilibrium ensemble in the high temperature limit.	In this particular case, and for a purely dissipative dynamics ($ \operatorname{H}_{t}=0 $) the insertion of the martingale process
	\begin{align}
		\mu_{t}=e^{2\,\mathscr{g}\,(t-\ti)}\,\lambda_{t}
		\nonumber
	\end{align}
	in the average maps the unraveling of the forward master equation (\ref{recovery:fme}) into that of the forward representation of the
	reversed dynamics (\ref{recovery:fbme}) according to (\ref{recovery:fbop}).

	\section{Recovery by embedding in a quantum channel}
	\label{sec:recqc}

\begin{figure}
		\centering
		\includegraphics{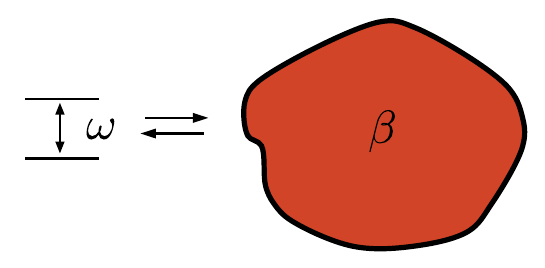}
		\caption{Stylized description of a qubit-bath interaction}
		\label{fig:qubit-bath}
	\end{figure}

\begin{figure}
		\centering
		\includegraphics{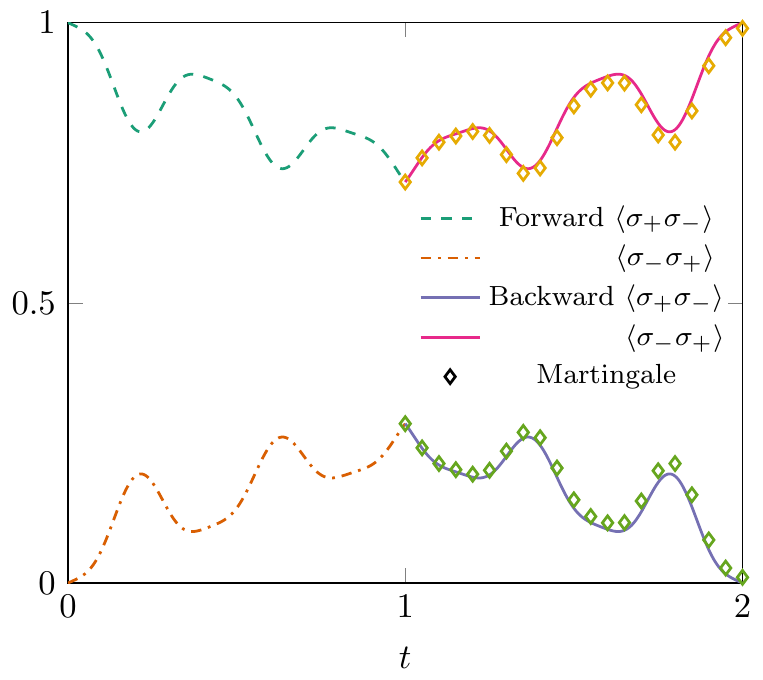}
		\caption{Example of a forward evolution described by the master equation \eqref{recqc:thermal}, with $g=0.1$ and $\beta=\omega=1$ and $\operatorname{H}_t=\sigma_3/2 +3\sin(15\,t)\sigma_1$ from $t=0$ to $t=1$. From $t=1$ to $t=2$ we recover the initial state. The full lines show the recovery using the embedding and performing an off-diagonal measurement. The diamonds show the recovery using the martingale. }
		\label{Fig:recovery}
	\end{figure}

	The embedding the completely bounded flow of (\ref{recovery:fbme}) in a completely positive map of section~\ref{sec:embedding} allows us to design an operational protocol to reverse a quantum evolution. The protocol consists of the following steps
		\begin{enumerate}[nosep, leftmargin=10pt, label=\roman*]
		\item we wish to recover $ \bm{\rho}_{\ti} $ given
		\begin{align}
			\bm{\rho}_{\tf}=\mathscr{P}_{\tf,\ti}(\bm{\rho}_{\ti}) 
			\nonumber
		\end{align}
		 where $ \mathscr{P}_{\tf,\ti} $ is a completely positive map solving for $ t\,\in \,[\ti,\tf] $  (\ref{recovery:fme}) or more generally (\ref{cb:LGKS});
		\item we couple the system to an ancillary qubit and define the tensor product state
		\begin{align}
			\bm{\gamma}_{\tf}=\frac{\operatorname{1}_{2}+\sigma_{1}}{2}\,\otimes\,\bm{\rho}_{\tf}
			\nonumber
		\end{align}
		\item $ \bm{\gamma}_{\tf} $ specifies the input of a quantum channel of the type (\ref{embedding:LGKS}) such that the off-diagonal corners of the flow generated by the channel satisfy 
		 \begin{align}
		 	\mathscr{B}_{t+\tf-\ti,\tf}=\mathscr{P}_{t,\ti}^{-1} \hspace{1.0cm}\forall\,t\,\in\,\left[\ti,\tf\right]
		 	\nonumber
		 \end{align}
		\item 	we obtain in output the initial value of the state operator by performing a measurement on the ancilla space according to
		\begin{align}
			\bm{\rho}_{\ti}=e^{\mathscr{g}\int_{\tf}^{2\tf-\ti}\mathrm{d}s\,\mathscr{c}_{s}}\operatorname{Tr}_{\mathfrak{1}}\left(\sigma_{1}\,\otimes\,\operatorname{1}_{\mathcal{H}} \bm{\gamma}_{2\tf-\ti}
			\right)
			\nonumber
		\end{align}
		where $ \operatorname{Tr}_{\mathfrak{1}} $ denotes the partial trace with respect to the first argument of the tensor product.
	\end{enumerate}
	In order to illustrate the protocol with an actual example, we consider the completely positive master equation
\begin{align}
	\label{recqc:thermal}
	\partial_{t}\bm{\chi}_{t}= -\,\imath\,[\operatorname{H}_t,\bm{\chi}_t] + g\,\operatorname{D}_{\sigma_{+}}\left(\bm{\chi}_{t}\right)
	+g\,e^{\beta\,\omega}\,\operatorname{D}_{\sigma_{-}}\left(\bm{\chi}_{t}\right)
\end{align}	    
which models the evolution of a driven qubit in contact with a thermal reservoir stylized in Fig. \ref{fig:qubit-bath}. The ladder operators
\begin{align}
	\sigma_{\pm}=\frac{\sigma_{1}\pm \imath\,\sigma_{2}}{2}
	\nonumber
\end{align}
readily satisfy (\ref{cb:povm}). The caption of Fig. ~\ref{Fig:recovery} specifies the parameters necessary to reproduce the numerics.

	We can also recover the initial state of the dynamics by unraveling the completely bounded dynamics describing the off-diagonal corner of the embedding flow with the influence martingale.	Points of the reverse trajectory computed by means of the influence martingale are marked by  diamonds in  Fig. ~\ref{Fig:recovery}.  From the numerical point of view, the use of the unraveling offers an advantage with respect to direct integration of the master equation only for systems with a sufficiently large number of states \cite{DoMG2022}. Convergence of the numerical simulations is in all cases guaranteed by standard results on the solutions of stochastic differential equations driven by counting processes \cite{PlBrLi2010}. We notice that, at least in principle, an operational implementation of the influence martingale algorithm is also possible. The implementation requires, however, post-selection by a classical apparatus. 	Namely, the protocol presumes storing the outcomes of continuous weak measurement records on a classical register. We suppose that each sequence of detected events in the record reconstructs a quantum trajectory. Hence each sequence must enter the average with a weigh determined by the corresponding realization of the influence martingale. This latter depends only on the measured detection rates which can be inferred from (\ref{unraveling:c}), (\ref{variance:optimal}).

	We conclude this section with some comments on the potential relevance of recovery protocol from the embedding  for quantum error correction.
	
	 One of the main challenges in quantum computing is to efficiently protect quantum memory from decoherence effects, construed as errors see e.g. \cite{WiMi2009,LiBr2013}. In particular, it is desirable to implement any quantum error correction process by means of a quantum circuit without invoking any classical apparatus and using only a few ancillary qubits.
	The recovery-by-embedding protocol only presumes completely positive operations and indirect measurement of an ancillary qubit. The full recovery the initial state of dynamics of a $ d $-state system only calls for the addition of one ancillary qubit. The protocol applies also in the case when the completely positive evolution in input is a particular solution of the canonical time local master equation. Over finite time intervals of time,
	the canonical time-local master equation provides a description of open quantum system dynamics equivalent to the Nakajima-Zwanzig equation.
	In this sense, the recovery-by-embedding protocol does not require modeling decoherence-induced errors by means of the Lindblad-Gorini-Kossakowski-Sudarshan equation that was criticized in \cite{AlLiZa2006}.

%	Finally, in our example, we illustrate recovery by reversing on the same footing the unitary evolution and decoherence channel and consider  constant in time decoherence rates. These choices are only for simplicity sake. 
	
	In summary, the influence martingale relates the recovery of the initial state operator evolved by a completely positive master equation to the unraveling of another completely positive master equation.

	\section{On the unraveling of non-canonical forms of the master equation}
	\label{sec:povm}
	
	We now turn to describe how to proceed in certain situations when  (\ref{cb:povm}) is not granted a priori.
	
	To start with, let us suppose that only some of the canonical decoherence rates $ \left\{ \mathscr{w}_{\mathscr{l}; t}\right\}_{\mathscr{l}=1}^{\mathscr{L}} $ in  (\ref{cb:LGKS}) are non-positive definite. In \cite{DoMG2022} we consider an example of this situation. It might be computationally convenient to construct an influence martingale enslaved only to the counting processes corresponding the \textquotedblleft non-positive definite\textquotedblright\ decoherence channels.	The apparent drawback is that we cannot expect that the corresponding operators satisfy the positive operator valued type condition  (\ref{cb:povm}).
	There is, however, a straightforward workaround to the problem. For simplicity of presentation, we illustrate the workaround surmising that in the master equation there are $ \mathscr{L}^{\prime} $ decoherence channel not satisfying  (\ref{cb:povm}), the extension to other related cases being straightforward.
	Indeed, we can always  include in the drift of the stochastic Schr\"odinger equation an additional operator $ \operatorname{L}_{0} $ such that
	\begin{align}
		\sum_{\mathscr{l}=0}^{\mathscr{L}^{\prime}}\operatorname{L}_{\mathscr{l}; t}\operatorname{L}_{\mathscr{l}; t}=\tilde{\mathscr{g}}^{\prime}\,\operatorname{1}_{\mathcal{H}}
		\nonumber
	\end{align}
	and, correspondingly, a counting process with increments $ \mathrm{d}\nu_{0; t} $ also satisfying the differential algebra (\ref{unraveling:counting}).	Next, we associate to the It\^o stochastic Schr\"odinger
	\begin{subequations}
		\label{povm:sse}
		\begin{align}
			&\label{povm:sse1}
			\mathrm{d}\bm{\psi}_{t}=\bm{g}_{t}\mathrm{d}t
			+\sum_{\mathscr{l}=0}^{\mathscr{L}^{\prime}}\mathrm{d}\nu_{\mathscr{l}; t}\left(\frac{\operatorname{L}_{\mathscr{l}; t}\bm{\psi}_{t}}{\left\|\operatorname{L}_{\mathscr{l}; t}\bm{\psi}_{t}\right\|}-\bm{\psi}_{t}\right)
			\\
			&\label{povm:sse2}
			\bm{g}_{t}=-\imath \operatorname{H}_{t}\bm{\psi}_{t}-\sum_{\mathscr{l}=0}^{\mathscr{L}^{\prime}}\mathscr{r}_{\mathscr{l}; t}\frac{\operatorname{L}_{\mathscr{l}; t}^{\dagger}\operatorname{L}-\left\|\operatorname{L}_{\mathscr{l}; t}\bm{\psi}_{t}\right\|^{2}}{2}\bm{\psi}_{t}
		\end{align}
	\end{subequations}
	the influence martingale equation
	\begin{align}
		&\mathrm{d}\mu_{t}=	\mu_{t}\left(\sum_{\mathscr{l}=1}^{\mathscr{L}^{\prime}}\left(\frac{\mathscr{w}_{\mathscr{l}; t}}{\mathscr{r}_{\mathscr{l}; t}}-1\right)\mathrm{d}\iota_{\mathscr{l}; t}-\mathrm{d}\iota_{0; t}\right)
		\nonumber\\
		&\mathrm{d}\iota_{\mathscr{l}; t}=\mathrm{d}\nu_{\mathscr{l}; t}-\mathscr{r}_{\mathscr{l}; t}\left\|\operatorname{L}_{\mathscr{l}; t}\bm{\psi}_{t}\right\|^{2}\mathrm{d}t
		\nonumber
	\end{align}
	In words, by enforcing the condition $ \mathscr{w}_{0; t}=0 $ the influence martingale suppresses paths solution of (\ref{povm:sse}) whenever a jump of the counting process $ \left\{ \nu_{0; t} \right\}_{t\,\geq\,\ti} $ occurs. It thus remain to verify that the unraveling conditions
	\begin{align}
		&	\mathscr{r}_{\mathscr{l}; t}=\mathscr{w}_{\mathscr{l}; t}+	\mathscr{c}_{t} \,\geq\,0
		\nonumber\\
		&	\mathscr{c}_{t} =2\,\max_{\mathscr{l}=1,\dots,\mathscr{L}^{\prime}}\left(-\mathscr{w}_{\mathscr{l}; t}\right)
		\nonumber
	\end{align}
	ensure that the drift (\ref{povm:sse2}) recovers the form (\ref{unraveling:drift}) hypothesized in the proof \cite{DoMG2022} of the unraveling applies.
	The dynamics (\ref{povm:sse}) preserve by construction the Bloch hyper-sphere as (\ref{unraveling:sse}) does. On the Bloch hyper-sphere the identity the chain of identities
	\begin{align}
		&\bm{g}_{t}=\sum_{\mathscr{l}=0}^{\mathscr{L}^{\prime}}(\mathscr{w}_{\mathscr{l}; t}+\mathscr{c}_{t})\frac{\operatorname{L}_{\mathscr{l}; t}^{\dagger}\operatorname{L}_{\mathscr{l}; t}-\left\|\operatorname{L}_{\mathscr{l}; t}\bm{\psi}_{t}\right\|^{2}}{2}\bm{\psi}_{t}
		\nonumber\\
		&	 =\mathscr{c}_{t}\sum_{\mathscr{l}=0}^{\mathscr{L}^{\prime}}\frac{\operatorname{L}_{\mathscr{l}; t}^{\dagger}\operatorname{L}_{\mathscr{l}; t}-\left\|\operatorname{L}_{\mathscr{l}; t}\bm{\psi}_{t}\right\|^{2}}{2}\bm{\psi}_{t}+
		\nonumber\\
		&		\sum_{\mathscr{l}=1}^{\mathscr{L}^{\prime}}\mathscr{w}_{\mathscr{l}; t}\frac{\operatorname{L}_{\mathscr{l}; t}^{\dagger}\operatorname{L}_{\mathscr{l}; t}-\left\|\operatorname{L}_{\mathscr{l}; t}\bm{\psi}_{t}\right\|^{2}}{2}\bm{\psi}_{t}
		\nonumber\\
		&=	\sum_{\mathscr{l}=1}^{\mathscr{L}^{\prime}}\mathscr{w}_{\mathscr{l}; t}\frac{\operatorname{L}_{\mathscr{l}; t}^{\dagger}\operatorname{L}_{\mathscr{l}; t}-\left\|\operatorname{L}_{\mathscr{l}; t}\bm{\psi}_{t}\right\|^{2}}{2}\bm{\psi}_{t}=\bm{f}_{t}
		\nonumber
	\end{align}
hold true. Hence, we recover (\ref{unraveling:drift}) and the proof of the unraveling by a completely positive stochastic state vector dynamics is complete. 
	
	Finally, we describe a physically relevant application of the workaround. As well known, the master equation (\ref{cb:LGKS}) is invariant if we replace Hamilton and decoherence operators with
	\begin{align}
		&		\tilde{\operatorname{L}}_{\mathscr{l}; t}=\operatorname{L}_{\mathscr{l}; t}+c_{\mathscr{l}; t}\,\operatorname{1}_{\mathcal{H}}
		\nonumber\\
		&		\tilde{\operatorname{H}}_{t}=\operatorname{H}_{t}-\frac{\imath}{2}	\sum_{\mathscr{l}=1}^{\mathscr{L}} \frac{\mathscr{w}_{\mathscr{l};t}}{2}
		\left(\bar{c}_{\mathscr{l}; t}\,\operatorname{L}_{\mathscr{l}; t}
		-
		c_{\mathscr{l}; t}\,\operatorname{L}_{\mathscr{l}; t}^{\dagger}
		\right)
		\nonumber
	\end{align}
	The unraveling is, however, not invariant under the transformation. In particular, the $ \left\{  \tilde{\operatorname{L}}_{\mathscr{l}; t}\right\}_{\mathscr{l}=1}^{\mathscr{L}} $ do not satisfy the positive operator valued measurement type condition (\ref{cb:povm}). 
	Nevertheless, the resulting stochastic Schr\"odinger equation models a measurement setup which in the completely positive case can be experimentally realized by homodyne detection \cite{WiMi2009}.
	
	\section{Discussion and outlook}
	
	The influence martingale \cite{DoMG2022} provides a general framework to unravel in quantum trajectories solutions of the canonical master equation to which any time-local open quantum system dynamics is always reducible \cite{HaCrLiAn2014}. Results such as those of \cite{AnCrHa2007,ChKo2010} prove the 
	equivalence of the  time-local and the time-non-local Nakajima-Zwanzig descriptions of a quantum open system dynamics, at least on finite time intervals \cite{vaWoLe2000}.  Hence, the influence martingale offers a general and in principle exact unraveling of any open system dynamics.
	In practice, however, application of the method generically relies on time convolutionless perturbation theory.     
	
	From the mathematical point of view, the unraveling consists in solving a system of $ d+1 $ ordinary stochastic It\^o differential equations. It is thus computationally equivalent to the unraveling of the completely positive master equation \cite{DaCaMo1992} because the statistics of state-vector dependent counting processes is usually reconstructed from Poisson processes by means of Girsanov's change of measure formula \cite{WisH1996}.
	
	The statistics generated by a stochastic Schr\"odinger equation models a weak measurement record. The physical interpretation of the measurement record requires the statistics to be non-anticipating (i.e. the present record cannot be affected by events in the future see  discussion e.g. in \cite{MeStLu2020}) and to establish a correspondence with an instrument i.e. a completely positive unital map see e.g. \cite{BaBe1991,BaHo1995,BaLu2005}.
	The influence martingale unraveling framework is by construction non-anticipating. Together, the first two main results of the present work imply that there are two \textquotedblleft natural\textquotedblright\ Hilbert spaces where we can relate a completely bounded flow to an instrument. The first is the original Hilbert space $ \mathcal{H} $, because the stochastic state vector evolution (\ref{unraveling:sse}) may equivalently be used for the unraveling of a completely bounded or completely positive master equation. The second is the embedding Hilbert space $ \mathbb{C}^{2}\,\otimes\,\mathcal{H} $, that can always be chosen to be the tensor product of the Hilbert space of the system with that of an ancillary qubit. The second avenue, embedding, is not new \cite{BrKaPe1999,BreH2004}. The embedding induced by the influence martingale may be regarded as, in some sense, minimal \cite{PaYuSu1985} and enjoys the property that diagonal an non-diagonal blocks of the completely positive flow are themselves flows of master equations in the original Hilbert space.	This latter property paves the way to applications to recovery of an initial state of a quantum open system evolution. 
	
	A criticism  \cite{AlLiZa2006} to continuous time error correction theory, is that modeling error build-up by the Lindblad-Gorini-Kossakowski-Sudarshan master equation may be inaccurate in realistic situations. We refer to \cite{McCaCrSaJa2020} for a recent quantitative appraisal of the criticism.  The recovery protocol that we propose does not necessarily require that the completely positive evolution to be reversed be the solution of a  Lindblad-Gorini-Kossakowski-Sudarshan master equation. It may well be a particular solution of the completely bounded master equation (see als o discussion in appendix (\ref{ap:flow})). The recovery-by-embedding protocol requires a detailed knowledge of the decoherence channels. This is however the case for any particular physical implementation of quantum computation (see e.g. supplementary information to \cite{ArBoBrMa2019}) and is therefore not a limitation specific of the protocol.
	
	In conclusion, the influence martingale unraveling framework provides a numerically efficient and conceptually ductile tool to analyze maps on operators and their \textquotedblleft matrix block\textquotedblright\ structure. A particularly interesting development is the quantum state recovery protocol that the influence martingale naturally brings about in view of potential applications to quantum error correction.

	\section{Acknowledgements} 
	The authors wish to thank Luca Peliti for insightful comments, and Jukka Pekola, Dmitry Golubev, Bayan Karimi and the Pico group at Aalto for many useful discussions and hospitality in their lab.
	
	\appendix
	
	\section{Memory effects as parametric dependence of the generator }
	\label{ap:flow}
	
	The derivation of (\ref{cb:LGKS}) in e.g. \cite{ChKo2010} surmises the existence of an instant of time $ \ti $ when the state operator of the microscopic bipartite system is the tensor product of the state operators of the constituents: system and environment. Upon tracing out the environment the system evolves from $ \ti $ according to a completely positive operator evolution map $ \Lambda $ solution of an integro-differential equation given by the Nakajiima-Zwanzig projection operator. This time-non-local equation is equivalent to a time-local master equation of the form 
	\begin{subequations}
		\label{flow:cp}
		\begin{align}
			&	\frac{\mathrm{d}\Lambda_{t,\ti}}{\mathrm{d} t}=\operatorname{L}_{t-\ti}(\Lambda_{t,\ti})
			\label{flow:me}
			\\
			&	\Lambda_{\ti,\ti}=\mathrm{Id}
			\label{flow:init}
		\end{align}	
	\end{subequations}
	in a time interval $ \left[\ti\,,\tf\right] $ where  $ \Lambda $ admits a continuous inverse. Upon assuming that the generator $ \operatorname{L}_{t} $ in $ [0\,,\tf-]\ti $ we may then regard (\ref{flow:cp}) as a special case of the equation 
	\begin{align}
		& \frac{\mathrm{d}\mathscr{B}_{t,s}}{\mathrm{d} t}=\operatorname{L}_{t-\ti}(\mathscr{B}_{t,s})
		\nonumber\\
		&\mathscr{B}_{s,s}=\mathrm{Id}
		\nonumber
	\end{align}
	defining a flow $\mathscr{B}_{s,s} $ for any $ s,t\in [0\,,\tf-\ti] $.  We immediately arrive at
	\begin{align}
		\Lambda_{t,\ti}=\mathscr{B}_{t-\ti,0}(\bm{\rho}_{\ti})
		\nonumber
	\end{align}
	Whilst $ \Lambda_{t,\ti} $ is by construction completely positive, the flow is generically completely bounded. To see this, we may consider 
	any $ v \,\leq\, t$ and use the group properties of the flow to establish the chain of identities
	\begin{align}
		\Lambda_{t,\ti}=\mathscr{B}_{t-\ti,v-\ti}\mathscr{B}_{v-\ti,0}=\mathscr{B}_{t-\ti,v-\ti}\Lambda_{v,\ti}
		\nonumber
	\end{align}
	We arrive at the identity 
	\begin{align}
		\mathscr{B}_{t-\ti,v-\ti}(\bm{\rho})=(\Lambda_{t,\ti}\Lambda_{v,\ti}^{-1})(\bm{\rho} )\hspace{1.0cm}\forall\,\bm{\rho}\in \mathcal{M}_{d}
		\nonumber
	\end{align}
	The inverse of completely positive operator evolution map is completely positive if and only if the map is unitary. Hence $ \mathscr{B}_{t-\ti,v-\ti} $
	is generically the composition of a completely positive with a completely bounded map and is therefore only completely bounded.

	\section{Proof of (\ref{cb:povm}) if the master equation is in canonical form}
	\label{ap:proof}
	
	If (\ref{cb:povm}) is in canonical form then $ \left\{  \operatorname{L}_{\mathscr{l}; t}\right\}_{\mathscr{l}=1}^{d^2-1} $ are related by a unitary transformation to the elements of an orthonormal basis of $ \mathcal{M}_{d}(\mathcal{H}) $ whose $\operatorname{L}_{d^{2}} $ is proportional to the identity \cite{BrPe2002}. Namely, upon writing the completeness relation in matrix components ($ \lhexbrace \operatorname{A} \rhexbrace_{l\,k}$ extricates the $ l,k $ entry from $ \operatorname{A} $) 
	\begin{align}
		\sum_{\mathscr{l}=1}^{d^{2}} \lhexbrace \operatorname{L}_{\mathscr{l}; t}\rhexbrace_{i\,j}\lhexbrace\operatorname{L}_{\mathscr{l}; t}^{\dagger}\rhexbrace_{l\,k}
		=\delta_{i\,k}\,\delta_{j\,l}
		\nonumber
	\end{align}
	we get
	\begin{align}
		\sum_{l=1}^{d}	\sum_{\mathscr{l}=1}^{d^{2}} \lhexbrace \operatorname{L}_{\mathscr{l}; t}\rhexbrace_{i\,l}\lhexbrace\operatorname{L}_{\mathscr{l}; t}^{\dagger}\rhexbrace_{l\,k}
		=\sum_{l=1}^{d}	\delta_{i\,k}\,\delta_{j\,j}=d\,\delta_{i\,k}
		\nonumber
	\end{align}
	which readily implies
	\begin{align}
		\sum_{\mathscr{l}=1}^{d^{2}-1}\operatorname{L}_{\mathscr{l}; t}\operatorname{L}_{\mathscr{l}; t}^{\dagger}=\left(d-\frac{1}{d}\right)\,\operatorname{1}_{\mathcal{H}}
		\nonumber
	\end{align}
	The result can also be read as a consequence of Schur's lemma applied to the generators of $ \mathfrak{su}(d) $.
	Furthermore the result is not affected by any unitary transformation of the basis elements.
	
	\section{Proof of the unraveling}
	\label{ap:Ito}
	
	It\^o lemma for stochastic processes with finite quadratic variation \cite{KleF2005} implies that
	\begin{align}
		\mathrm{d}\left(\mu_{t}\bm{\psi}_{t}\bm{\psi}_{t}^{\dagger}\right)=
		(\mathrm{d}\mu_{t})\bm{\psi}_{t}\bm{\psi}_{t}^{\dagger}+(\mu_{t}+\mathrm{d}\mu_{t})\mathrm{d}\left(\bm{\psi}_{t}\bm{\psi}_{t}^{\dagger}\right)
		\nonumber
	\end{align}
	and
	\begin{align}
		\mathrm{d}\left(\bm{\psi}_{t}\bm{\psi}_{t}^{\dagger}\right)=(\mathrm{d}\bm{\psi}_{t})\bm{\psi}_{t}^{\dagger}+\bm{\psi}_{t}\mathrm{d}\bm{\psi}_{t}^{\dagger}+(\mathrm{d}\bm{\psi}_{t})\mathrm{d}\bm{\psi}_{t}^{\dagger}
		\nonumber
	\end{align}
	The explicit expressions of the differentials (\ref{unraveling:sse1}), (\ref{martingale})  and the telescopic property of expectation values in the It\^o prescription e.g.
	\begin{align}
		\operatorname{E}\left(\mathrm{d}\nu_{\mathscr{l}; t} \bm{\psi}_{t}\bm{\psi}_{t}^{\dagger}\right)=
		\operatorname{E}\left(\operatorname{E}\left(\mathrm{d}\nu_{\mathscr{l}; t}\big{|} \mathcal{F}_{t}\right)\bm{\psi}_{t}\bm{\psi}_{t}^{\dagger}\right)
		\nonumber
	\end{align}
	which immediately imply
	\begin{align}
		\operatorname{E}\left(\mathrm{d}\mu_{ t} \bm{\psi}_{t}\bm{\psi}_{t}^{\dagger}\right)=\operatorname{E}\left(\mathrm{d}\iota_{\mathscr{l}; t} \bm{\psi}_{t}\bm{\psi}_{t}^{\dagger}\right)=0\hspace{0.5cm}\forall\,\mathscr{l}
		\nonumber
	\end{align}
	prove that (\ref{unraveling:main}) satisfies (\ref{cb:LGKS}).
	
	\section{Commutant representation}
	\label{ap:commutant}
	
	We recall that the adverb completely in reference to a property enjoyed by a linear map $ \mathscr{B} $ on $ \mathcal{B}(\mathcal{H}) $ indicates that that the same property is enjoyed by the extension
	\begin{align}
		\mathrm{Id}_{k}\,\otimes\,\mathscr{B}\colon \begin{bmatrix}
			\operatorname{O}_{1\,1} & \dots & 	\operatorname{O}_{1\,k}\\ \vdots &   \vdots &   \vdots \\\operatorname{O}_{k\,1} & \dots &\operatorname{O}_{k\,k}
		\end{bmatrix}\mapsto
		\begin{bmatrix}
			\mathscr{B}(\operatorname{O}_{1\,1}) & \dots & \mathscr{B}(\operatorname{O}_{1\,k})\\ \vdots &   \vdots &   \vdots \\
			\mathscr{B}(\operatorname{O}_{k\,1}) & \dots & \mathscr{B}(\operatorname{O}_{k\,k})
		\end{bmatrix}
		\nonumber
	\end{align}
	for any $ k\in \mathbb{N} $ and $ \operatorname{O}_{i\,j} $'s in $ \mathcal{B}(\mathcal{H}) $. 
	
	We now summarize some of the results of \cite{PaYuSu1985} about the representation of completely bounded maps 
	also drawing from the pedagogic presentations \cite{PauV2007,JoKrPa2009}.
	Restricting for simplicity the discussion to finite dimensional spaces, the commutant representation of a completely bounded map $ \mathscr{B}\colon \mathcal{M}_{d} \mapsto\mathcal{M}_{\tilde{d}}$ consists in the fact that there exists a collection of $ \tilde{d}\,\times\,d $ rectangular matrices $\operatorname{A}_{\mathscr{i}} $ $ \mathscr{i}=1,\dots,\mathscr{m}\,\leq\,d \,\tilde{d} $ satisfying the positive operator value measurement condition
	\begin{align}
		\sum_{\mathscr{i}=1}^{\mathscr{m}}\operatorname{A}_{\mathscr{i}}\operatorname{A}_{\mathscr{i}}^{\dagger}=\operatorname{1}_{\tilde{d}}
		\nonumber
	\end{align}
	and a matrix $ \operatorname{T}\in \mathcal{M}_{\mathscr{m}} $ such that for any $ \operatorname{X}\in  \mathcal{M}_{d} $ we can write
	\begin{align}
		\mathscr{B}(\operatorname{X})=\begin{bmatrix} \operatorname{A}_{1} & \dots & \operatorname{A}_{\mathscr{m}} \end{bmatrix}  (\operatorname{T} \otimes \operatorname{1}_{\mathcal{H}} ) ( \operatorname{1}_{\mathscr{m}} \otimes \operatorname{X})\begin{bmatrix} \operatorname{A}_{1}^{\dagger} \\ \vdots \\ \operatorname{A}_{\mathscr{m}}^{\dagger} \end{bmatrix} 
		\label{commutant:repr}
	\end{align}
	The name \textquotedblleft commutant\textquotedblright\ stems from the observation
	\begin{align}
		(\operatorname{T} \otimes \operatorname{1}_{\mathcal{H}} ) ( \operatorname{1}_{\mathscr{m}} \otimes \operatorname{X})=( \operatorname{1}_{\mathscr{m}} \otimes \operatorname{X})(\operatorname{T} \otimes \operatorname{1}_{\mathcal{H}} ) 
		\nonumber
	\end{align}
	In particular if $ \operatorname{T}=\operatorname{1}_{\mathscr{m}} $ we recover the Choi-Stinespring representation of a completely positive map. It is then convenient to write (\ref{commutant:repr}) in the Krauss operator product form
	\begin{align}
		\mathscr{B}(\operatorname{X})=\sum_{\mathscr{i},\mathscr{j}=1}^{\mathscr{m}}\operatorname{T}_{\mathscr{i},\mathscr{j}}\operatorname{A}_{\mathscr{i}}\operatorname{X}\operatorname{A}_{\mathscr{j}}^{\dagger}
		\nonumber
	\end{align}
	A consequence \cite{PaYuSu1985} of the existence  of the commutant representation is that the embedding linear map $ \mathscr{E}\colon \mathbb{C}^{2} \,\otimes\,\mathcal{M}_{d} \mapsto \mathbb{C}^{2} \,\otimes\,\mathcal{M}_{\tilde{d}}$ defined by
	\begin{align}
		&	\mathscr{E}\left(\begin{bmatrix}
			\operatorname{Y}	& \operatorname{X} \\\operatorname{W}	& \operatorname{Z} 
		\end{bmatrix}\right)
		=
		\nonumber\\
		&\begin{bmatrix}
			\sum_{\mathscr{i}}^{\mathscr{m}}\operatorname{A}_{\mathscr{i}}\operatorname{Y}\operatorname{A}_{\mathscr{i}}^{\dagger}&	\sum_{\mathscr{i},\mathscr{j}=1}^{\mathscr{m}}\operatorname{T}_{\mathscr{i},\mathscr{j}}\operatorname{A}_{\mathscr{i}}\operatorname{X}\operatorname{A}_{\mathscr{j}}^{\dagger}\\  \sum_{\mathscr{i},\mathscr{j}=1}^{\mathscr{m}}\bar{\operatorname{T}}_{\mathscr{i},\mathscr{j}}\operatorname{A}_{\mathscr{i}}^{\dagger}\operatorname{W}\operatorname{A}_{\mathscr{j}} &
			\sum_{\mathscr{i}}^{\mathscr{m}}\operatorname{A}_{\mathscr{i}}\operatorname{Y}\operatorname{A}_{\mathscr{i}}^{\dagger}
		\end{bmatrix}
		\nonumber
	\end{align}
	is completely positive if and only if the $ 2\,\mathscr{m}\,d\,\times\, 2\,\mathscr{m}\,d$ squared matrix
	\begin{align}
		\operatorname{M}=\begin{bmatrix}
			\operatorname{1}_{\mathscr{n}} \otimes \operatorname{1}_{\mathcal{H}} &\operatorname{T} \otimes \operatorname{1}_{\mathcal{H}} 	\\  
			\operatorname{T}^{\dagger} \otimes \operatorname{1}_{\mathcal{H}} 	 & \operatorname{1}_{\mathscr{n}} \otimes \operatorname{1}_{\mathcal{H}} 
		\end{bmatrix}
		\nonumber
	\end{align}
	is positive definite. The completely positive map
	\begin{align}
		\mathscr{P}(Y)=\sum_{\mathscr{i}}^{\mathscr{m}}\operatorname{A}_{\mathscr{i}}\operatorname{Y}\operatorname{A}_{\mathscr{i}}^{\dagger}
		\nonumber
	\end{align}
	appearing in the embedding map is called the \textquotedblleft associated\textquotedblright\ completely positive map. Finally, \cite{PaYuSu1985}
	addresses the problem of the uniqueness of the representation and proves that the commutant representation is unique up to similarity under the same conditions as the Choi-Stinespring representation in the completely positive case \cite{PauV2003}.
	
	\bibliography{pairing}{} % the path must be given in reference to the compilation directory
	\bibliographystyle{apsrev4-2}
\end{document}